\documentclass [10pt,a4paper]{article}
\usepackage{jheppub}

\usepackage[latin1]{inputenc}
\usepackage{graphicx}
\usepackage{color}
\usepackage{amsfonts,amsthm,amssymb}
\usepackage{bm,bbm}
\usepackage[OT2,OT1]{fontenc}
\usepackage{nicefrac}
\newcommand\cyr{%
\renewcommand\rmdefault{wncyr}%
\renewcommand\sfdefault{wncyss}%
\renewcommand\encodingdefault{OT2}%
\normalfont
\selectfont}
\DeclareTextFontCommand{\textcyr}{\cyr}

\def\beq{\begin{equation}}
\def\eeq{\end{equation}}
\newcommand{\be}{\begin{eqnarray}}
\newcommand{\ee}{\end{eqnarray}}

\renewcommand{\texttt}{{}}

\def\bs{\begin{subequations}}
\def\es{\end{subequations}}

\newcommand{\tia}[1]{}

\newcommand{\bea}{\begin{eqnarray}}
\newcommand{\eea}{\end{eqnarray}}
\newcommand{\beas}{\begin{eqnarray*}}
\newcommand{\eeas}{\end{eqnarray*}}
\newcommand{\bal}{\begin{aligned}}
\newcommand{\eal}{\end{aligned}}

\def\({\left(}
\def\){\right)}

\newcommand{\half}{\frac{1}{2}}

\begin{document}

\title{Exact solutions and spacetime singularities\\ in nonlocal gravity}

\author{Yao-Dong Li, Leonardo Modesto, and Les\l aw Rachwa\l}

\affiliation{Department of Physics \& Center for Field Theory and Particle Physics, \\
Fudan University, 200433 Shanghai, China}

\emailAdd{ydli12@fudan.edu.cn}
\emailAdd{lmodesto@fudan.edu.cn}
\emailAdd{rachwal@fudan.edu.cn}

\abstract{
We give here a list of exact classical solutions of a large class of weakly nonlocal theories of gravity, which are unitary and super-renormalizable (or finite) at quantum level. It is explicitly shown that flat and Ricci-flat spacetimes as well as maximally symmetric manifolds are exact solutions of the equation of motion. Therefore, well-known physical spacetimes like Schwarzschild, Kerr, (Anti-) de Sitter serve as solutions for standard matter content. In dimension higher than four we can also have Anti-de Sitter solutions in the presence of positive cosmological constant. We pedagogically show how to obtain these exact solutions. Furthermore, for another version of the theory, written in the Weyl basis, Friedmann-Robertson-Walker (FRW) spacetimes are also exact solutions, when the matter content is given by conformal matter (radiation). We also comment on the presence of singularities and possible resolution of them in finite and conformally invariant theories. ``Delocalization" is proposed as a way to solve the black hole singularity problem. In order to solve the problem of cosmological singularities it seems crucial to have a conformally invariant or asymptotically free quantum gravitational theory.}

\maketitle



\date{\small April 7, 2014}


\section{Introduction}

In this paper we study a class of exact solutions of nonlocal gravity theories and discuss the issues related to the presence of classical singularities \cite{Hawking:1969sw}. The reasons for studying classical solutions of this nonlocal theory motivated by quantum theory \cite{modesto,modestoLeslaw, Briscese:2013lna, Krasnikov, Tombo, M3, M4, Khoury:2006fg, Mtheory, Modesto:2013jea} are at least three-fold. 

First, it was shown that this proposal for an action of quantum gravity solves its all burning problems, like unitarity and renormalizability. Moreover, we believe that due to the presence of nonlocality in the classical (bare) action, this action ansatz is very close to the true quantum effective action in quantum gravity, which is the result of resummation of all perturbative loop contributions. Since tree-level correlations functions derived from the quantum effective action capture all quantum physics effects, it is enough to study classical solutions of the effective action. This is equivalent to the investigation of full quantum solutions of the Heisenberg equation of motion. The last ones are very difficult to obtain based on quantized classical theory, hence our method can be very useful in determining the character of genuine quantum corrections to the classical solutions of general relativity found within Einstein theory. 

The second reason of our study is of more technical character. To perform the quantization of some classical theory first it is required to find its vacua being classical exact solutions. Only upon them we can quantize the system of fields and end up with quantum field theory. At the end we must check for consistency and whether these vacua are solutions also in the quantum version of the theory. Therefore, the knowledge of classical exact solutions is of great importance for study quantization on different backgrounds. 

The third reason is dedicated to the study of singularities. Spacetime singularities are the crucial issues that plague Einstein classical theory of gravity \cite{Hawking:1969sw}. It is believed that quantum gravity will solve such problems, but this is not so automatic as will be pointed out in this paper. We can start from a new action principle for gravity that is super-renormalizable or even finite at quantum level, and still have a large class of singular mathematical solutions. If the theory is UV-finite we can hope that such spacetimes will be thrown out of the set of solutions of the quantum action, but only the knowledge of the full effective action will confirm such a conjecture. However, if our new classical theory is weakly nonlocal a new mechanism is at work, namely a ``delocalization" of the source: any ``Dirac delta-like" $p$-brane source will be effectively spread in the whole spacetime, therefore endowing it with an infinitesimal hair, slightly violating the no-hair theorem. Moreover, it is well known that $\mathcal{N}=4$ super-Yang-Mills theory is finite at quantum level, but the gauge potential (Coulomb potential) is singular at $r=0$, while in super-renormalizable or finite gravity theories 
\cite{Stelle, Shapirobook, shapiro3, HigherDG, HigherDG0,  modesto} the Newtonian potential approaches a constant value at short distances \cite{StelleCS, Frolov, ModestoMoffatNico,Tiberio, BM, BambiMalaModesto2, BambiMalaModesto, calcagnimodesto, Yiwei, koshe1}. Once more it is not so obvious whether quantum mechanics may help or not in resolving and smearing out the singularities. The gauge theory example mentioned above suggests the possibility that it is the classical nonlocality (or classical locality with higher derivatives  \cite{Tiberio}) that makes the potential singularity-free. 

In this paper we first remind the reader of the class of super-renormalizable or finite theories of gravity, after which we list and study a large set of classical exact solutions common to any local or weakly nonlocal higher derivative theory. We will show in particular that all vacuum solutions (Schwarzschild, Kerr, etc.), Einstein manifolds $R_{\mu\nu} = {\rm const} \cdot g_{\mu\nu}$ and the Friedman-Robertson-Walker (FRW) cosmological models with radiation (and/or cosmological constant) are exact solutions of our candidate finite theory and of a larger class of higher derivative theories. In all above cases we discuss the problem of singularities, whether they are avoided in higher derivative nonlocal theories or whether they are still present as in two-derivative Einstein theory. We propose two mechanisms to exclude remaining solutions with singularities from the set of physically admissible spacetimes. For the case of Ricci-flat solutions we propose ``delocalization" of the matter source, while for cosmological solutions we invoke conformal invariance to eventually wash out the Big Bang singularity. Moreover, for the purpose of finding cosmological solutions we describe the theory written in the Weyl basis and discuss properties thereof. Later we discuss an issue of multi-scale black holes in vacuum, which are also possible solutions of the theory in the presence of a new mass scale. Finally after a little summary of what has been learned and how it is possible to solve the problem of singularities, we add to the paper two appendices, where we explicitly, with all details and indices, derive EOM of the theory in the case of one and two form factors respectively.

Let us begin with recalling a general class of $D$-dimensional weakly nonlocal theories quadratic in the Riemann, Ricci and scalar curvature tensors, 
\cite{modesto,modestoLeslaw, Briscese:2013lna, Krasnikov, Tombo, BM, M3, M4, Khoury:2006fg, Mtheory, Modesto:2013jea},
\be
&& \hspace{-.8cm} \mathcal{L}_{\rm g} = -  2 \kappa_{D}^{-2} \, \sqrt{|g|} 
\Big[\Lambda_{\rm cc} + {\bf R} + {\bf R} \, \gamma_0(\Box) {\bf R } + {\bf Ric} \, \gamma_2(\Box) {\bf Ric} + {\bf Riem}  \, \gamma_4(\Box) {\bf Riem} + {\bf {V}} \, \Big]. 
\label{gravity}
\ee
The theories above consist of a weakly nonlocal kinetic operator and a curvature potential $\mathbf{V}$ which is chosen to be local here,
\be
{\bf V} = \mathcal{R}^3 + \dots  + \!\!\! \sum_{k=3}^{\gamma +{\rm N}+2} \sum_i s_{k,i} \, \left(  \nabla^{2 (\gamma + {\rm N}+2 -k )} \, {\cal R}^k \right)_i ,  
\label{K0}
\ee
 where the operators in the last set are called {\bf killers} because they are crucial in making the theory finite in any dimension. (They kill the beta functions as explained thoroughly in \cite{modestoLeslaw, universality, Modesto:2015foa}.)
 In (\ref{gravity}) $\Lambda_{\rm cc}$ is a cosmological constant term, $s_{k,i}$ in \eqref{K0} are numerical coefficients, and the tensorial structures above have been neglected\footnote{{\em Definitions ---} The metric tensor $g_{\mu \nu}$ has signature $(- + \dots +)$ and the curvature tensors are defined as follows: $R^{\mu}_{\nu \rho \sigma} = - \partial_{\sigma} \Gamma^{\mu}_{\nu \rho} + \dots $, $R_{\mu \nu} = R^{\rho}_{\mu  \rho \nu}$,  
$R = g^{\mu \nu} R_{\mu \nu}$. With symbol ${\cal R}$ we generally denote one of the above curvature tensors.}. Moreover, $\Box = g^{\mu\nu} \nabla_{\mu} \nabla_{\nu}$ is the covariant box operator, the integer parameter $\gamma$ and form factors $\gamma_0, \gamma_2, \gamma_4$  will be defined shortly after. The capital~$\rm{N}$ is defined to be the following function of the spacetime dimension $D$: $2 \mathrm{N} + 4 = D$ for even $D$ and $2 \mathrm{N} + 4 = D+1$ for odd $D$. For the minimal theory ($\gamma_4 = 0$) unitarity requires \cite{HigherDG}
\be
\label{formfactor}
\gamma_0(\Box) = - \frac{\gamma_2(\Box)}{2} =  - \frac{e^{H(-\Box_{\Lambda})} -1}{2 \, \Box} \,,
\ee
where $\Box_\Lambda\equiv\Box/\Lambda^2$ and $\Lambda$ is an invariant fundamental mass scale of the theory.

A universal exponential form factor $\exp H\left(- \Box_{\Lambda}\right)=\exp H\left(z\right)$ compatible with the guiding principles of quantum field theory pointed out by Tomboulis is \cite{Tombo}:
\be
\label{Tombconstruction}
\hspace{-1.5cm}
&& e^{H(z)}= e^{\frac{a}{2} \left[ \Gamma \left(0, p(z)^2 \right)+\gamma_E  + \log \left( p(z)^2 \right) \right] } \\
&&\hspace{0.9cm} = e^{a \frac{\gamma_E}{2}} \,
\sqrt{ p(z)^{2 a}} \left\{ 1+ \left[ \frac{a \, e^{-p(z)^2}}{2 \,  p(z)^2} \left( 1 + O \left(   \frac{1}{p(z)^2} \! \right)   \! \right) + O \left(e^{-2 p(z)^2} \right)  \right] \right\}  , 
\label{Tomboulis}
\ee
where the last equality is correct only on the real axis ($\gamma_E \approx 0.577216$ is the Euler-Mascheroni constant and $\Gamma(0,z) = \int_z^{+ \infty}  d t \, e^{-t} /t$ is the incomplete gamma function with its first argument vanishing). The  polynomial $p(z)$ of degree $\gamma +\mathrm{N}+1$ is such that $p(0)=0$, which gives the correct low energy limit of our theory (coinciding with Einstein gravity) and $a$ is a positive integer parameter. The entire function has in UV asymptotics ($|z|\gg1$) polynomial behaviour $z^{a(\gamma + {\rm N} +1)}$ in conical regions around the real axis with angular opening 
$\Theta= \pi/(4 (\gamma + \mathrm{N} + 1))$. For $\gamma =0$ we have the maximal conical region with an opening angle $\Theta = \pi/4$ in $D = 4$ for all $a$.

The explicit field equation in the theory \eqref{gravity} for the case $\mathbf{V}=0$ is given in the appendix.

\section{Flat and Ricci-flat spacetimes}

In the following three sections we infer about a large class of exact solutions in local or weakly nonlocal higher derivative theories
without selecting out any specific action.
The building blocks with which we build the most general theory are made of covariant derivative of the curvature scalar, Ricci, and Riemann tensor,
\be 
{\bf \nabla^m R^n} \, , \,\,\,\, {\bf  \nabla^m Ric^n} \, , \,\,\,\, {\bf \nabla^m Riem^n} \, , \,\,\,\, {\rm for} \,\, {\bf n,m}\geqslant 0.
\ee
Our analysis will be general for any local or weakly nonlocal higher derivative theory, but we will often refer 
to the class of unitary and super-renormalizable theories introduced in the previous section. We start with vacuum solutions of the theory, consisting of flat and Ricci-flat spacetimes in the simplest version of the theory similar to standard Einstein-Hilbert gravity. In the course we will state the conditions for which theories and for which matter content these spacetimes are exact solutions.

\subsection{The Minkowski spacetime} 

Flat spacetime $g_{\mu\nu} = \eta_{\mu\nu}$ is a solution in any generally covariant theory of gravity provided that cosmological constant $\Lambda_{\rm cc}	$ is zero and without any other matter source. The reason is simple: we get contributions containing curvature tensors from every term in the gravitational part of the action (except the cosmological constant), which will vanish when evaluated on the flat spacetime. Actually, in the EOM for the gravitational field, written in a fully covariant form, the derivatives act on the gravitational curvature tensors as a result of one variation of the operators in the action (the derivatives will always be there when we have dynamics in the minimal number of two and always even in number because we have a bosonic field) :
\be
	\mathbf{G} + \nabla^n \mathbf{R} + \Lambda_{\rm cc} \mathbf{g} = 0, \,{\mbox{ where $n$ is a positive integer.}}
\ee
The term resulted from cosmological constant is proportional to the spacetime metric, with proportionality constant independent from the spacetime point.
Now, when evaluating all other terms in the covariant EOM on the flat spacetime all gravitational curvatures vanish, so from all terms containing any curvature we get zero contribution. The EOM is satisfied only if the presence of the cosmological constant is accompanied by the matter energy tensor, which has exactly the same form as that of the cosmological constant, hence proportional to the metric and the coefficient is exactly minus the value of the cosmological constant. To be more precise, the flat spacetime is a solution, only if
\be \Lambda_{\rm cc} g^{\mu\nu}+T^{\mu\nu}=0 \,\,\, {\mbox{ or in tensorial compact notation}} \,\,\, 
\Lambda_{\rm cc} {\bf g} + {\bf T} = 0 \label{cc_matter_relation}
,
\ee 
where $T^{\mu\nu}$ is the energy tensor from the full matter sector.

\subsection{Ricci-flat spacetimes}
The next simple generalizations are the Ricci-flat spacetimes as solutions of the gravitational EOM in vacuum, so without any matter energy tensor and without cosmological constant term.\footnote{In principle one does not have to assume that both cosmological constant and matter energy tensor vanish; however, as long as \eqref{cc_matter_relation} is satisfied, there is no difference between the two cases at the level of EOM.} On Ricci-flat manifolds the Ricci tensor 
{\bf Ric}
and the scalar curvature {\bf R} vanish everywhere, but the metric is not flat because the Riemann tensor
{\bf Riem} is non-vanishing. 

\subsubsection{General criteria for Ricci-flatness ansatz}
\label{critRicflat}

We shall give an analysis valid for the general Ricci-flat metrics.

Knowing the characteristic feature of these types of manifolds we can easily specify the conditions when the full gravitational EOM is satisfied on these manifolds. We recall some facts about the first variations of the covariant building blocks (\ref{building-block})--(\ref{var_tensorial_box}) of the gravitational actions like curvature tensors and covariant derivatives acting on them. The most general (on the general curved background) first variation of the Ricci tensor \eqref{var_Ric} and the scalar \eqref{var_Rs} does not contain Riemann curvature tensor; it may contain only Ricci tensor as a gravitational background curvature. The case for the variation of the Riemann tensor is of course different, because it does contain its own background value.

Regarding the first variations of differential operators \eqref{var_scalar_box} and \eqref{var_tensorial_box} (like the covariant Beltrami-Laplace box operator or the covariant derivatives), they typically do not contain any curvature tensor of the background, because they were not present there even before the variation was taken. These facts are necessary ingredients to assess the general form of the EOM in any generally covariant gravitational theory.

We will concentrate mostly on the presence of different curvature tensors in this equation.
Recall that this equation results from taking the first variation of the generally covariant gravitational action. It is easily understood that on Ricci-flat manifolds the presence of the Ricci tensor or the scalar in a term in the EOM makes it vanish when the equation is evaluated on the Ricci-flat metric ansatz (it does not make any difference whether these Ricci curvatures are differentiated or not). Every time we meet at least one power of Ricci curvature tensor (where by this we mean Ricci scalar or tensor) in a term, we can easily forget about this term in checking the validity of the Ricci-flat ansatz for the EOM in some particular theory.

The case that deserves further investigation is only when there is a full Riemann tensor involved. The latter generally does not vanish on Ricci-flat manifold and we do not have any guaranty to throw away such term in the EOM.
Typically its contribution will be non-zero and it could spoil our ansatz for Ricci-flat metrics as the solutions in this particular theory, but not necessarily. The problem comes when in some terms of the EOM there are no other Ricci curvature tensors, but only Riemann tensors or covariant derivatives acting on them. Such terms will contribute to the EOM, and in their presence the Ricci-flat ansatz might no longer be valid.
The analysis in this situation is more involved and may require taking into total account other terms in the EOM with Riemann curvature tensors; it maybe even necessary to check the behaviour of terms with {\bf Riem} on the particular form of the metric (still satisfying the Ricci-flatness ansatz). 

However, we are able to derive a more useful criterion without the need to explicitly compute the first variation and getting the covariant EOM. We can perform the analysis completely on the level of covariant gravitational action. Due to the reasons presented above only the variation of {\bf Riem} gives rise to the background Riemann curvature tensor. Hence the same analysis can be performed on the level of action. If in its terms we meet at least two Ricci curvatures
${\bf Ric}^2$, 
then the contribution of this term to the EOM evaluated on the Ricci-flat ansatz will certainly vanish. The criterion fails only when there is explicit usage of the Riemann tensor in the construction of building blocks for the covariant action or if there is precisely one Ricci curvature contracted with covariant derivatives or other Riemann tensors. 
Only Riemann tensors and covariant derivatives acting on them in some operator term 
(if not accompanied by two or more Ricci curvatures) requires detailed analysis, which must be done on the level of the EOM, so in such circumstances it is necessary to take variation of this term to be able to conclude about the validity of the ansatz. The simplest examples of such problematic terms in the action, written with full index structure, are 
\be 
R_{\mu\nu\rho\sigma}R^{\mu\nu\rho\sigma} \, ,  \,\,\,\, R_{\mu\nu}R^\mu{}_{\rho\sigma\tau}R^{\nu\rho\sigma\tau}
\, , \,\,\,  \nabla_\mu\nabla_\nu\nabla_\rho\nabla_\sigma R^{\mu\nu\rho\sigma}.
\ee  

Another source of hope for the validity of the Ricci-flat ansatz may come from the fact that in the procedure of getting the EOM from the action we throw away total derivatives appearing in the variation of the action and moreover, we can always integrate by parts. This may reduce the number of terms with full Riemann tensors in the final EOM.
Without deriving the EOM in its full form (which is fixed and unambiguous), it is possible to see at the level of action that some special combinations of terms with $\mathbf{Riem}$ cannot contribute, for they themselves or their variations are total derivatives. Examples include Gauss-Bonnet terms and terms of higher order in $\mathbf{Riem}$ in higher dimensional spaces \cite{'tHooft:1974bx}. We will comment more on this issue in a later subsection.

\subsubsection{Ambiguity in power counting of curvatures} 
\label{ambiguity}

One issue needs also to be clarified regarding the criterion put above, namely it is known that counting of curvature tensors in a generally covariant expression with higher derivatives is {\em ambiguous}. We can easily change their number by doing the commutation of covariant derivatives, at the price of getting more Riemann tensors or Ricci tensors, a simple example being
\be
	\nabla_1 \nabla_2 \mathbf{T} = \nabla_2 \nabla_1 \mathbf{T} + \mathbf{Riem} \cdot \mathbf{T}.
\ee
To use our criterion above unambiguously we must specify what to do with this ambiguity of the order of covariant derivatives on tensors. We may think that a resolution is to declare that all covariant derivatives in the expression 
are ``normal ordered'' (or symmetrized in all tensorial indices) in a convenient way as if they commuted and then compute the difference between this expression (with covariant derivatives put in whatever order) and the actual expression. In the expression with flat covariant derivatives their order will be completely immaterial, so we may expect some number of cancellations between different terms and hence the result should be shorter. In the difference we will produce curvature tensors resulting from the commutation (this procedure may produce finitely many terms for theories with a finite number of derivatives). The result is an expression with ordered sequences of covariant derivatives in each term and with a bunch of terms higher in gravitational curvatures. This final result is unambiguous, but it is in the very opposite direction to the analysis we presented above. Due to these {\em unambiguization} procedure
of the covariant expression new curvature tensors are often produced, and it may happen (even many times) that Riemann tensors are produced and hence our analysis needs a refinement. Therefore, it is much better for us not to commute these unordered covariant derivatives. The conclusion is that, if the EOM contains unordered sequences of these derivatives, but not terms with {\bf Riem} alone, then the contribution to the EOM will vanish on the ansatz. Commuting the derivatives may darken out the analysis, because then the Riemann tensors will be produced. Actually the direction of our analysis is rather opposite, because if we can show that in a term the Riemann tensor is actually the result of commutation of derivatives, then we can come back to the original form of writing it (with two derivatives only) and therefore reduce the number of existing Riemann tensors in the expression. Later we may more easily conclude, that this term does not give any non-vanishing contribution to the EOM on the ansatz.

The previous remark with ambiguity in counting the powers of curvatures is valid on the level of action, as well as on the level of EOM, because it is possible to commute covariant derivatives or uncommute the Riemann tensor in any valid tensorial expression. In order to perform uncommutation (like reversing the unambiguization procedure described above) of the Riemann curvature tensor from a covariant expression on any level (action or EOM) one must be careful and perform it not on just one term, but on the special combination of terms 
resulting from the same commutation of derivatives (typically commutation of derivatives is done when they act on a tensor, and here we have at our disposal always only tensors with even valence, hence more than one term with higher derivatives will be produced). 
This is the reason why we have to look at the combination of terms, if we hope to write Riemann tensors appearing in them as a result of commutation of covariant derivatives.  

In the whole discussion above we can equally well substitute all the arguments about the Riemann tensor with the Weyl tensor or any other tensor of the gravitational curvature with four indices, with the help of changes of basis. 
Here we will concentrate on all possible terms quadratic in curvature (having in mind the ambiguity of such counting) in spacetime dimension $D=4$. At the beginning we can start with terms that are with organized structure of derivatives, namely they appear only in the form of covariant box operators. Moreover, exploiting gravitational Bianchi identities we can always reduce the quadratic action to such forms up to covariant terms higher than quadratic in curvatures. Actually in any dimension there are only 3 independent elements of the basis of such terms with fixed number of derivatives in boxes. They have the structure of two identical tensors, whose indices are contracted across,
\be
	R \Box^n R, \,\, R_{\mu\nu} \Box^n R^{\mu\nu}, \,\, R_{\mu\nu\alpha\beta}\Box^n R^{\mu\nu\alpha\beta}, \,\, C_{\mu\nu\alpha\beta}\Box^n C^{\mu\nu\alpha\beta}.
\ee
The box operators are inserted in the middle, so they all act on the right curvature tensor. We can always make a choice of basis of 3 elements out of this ``linearly dependent'' set of 4 by performing a sufficient number of direct substitutions. For convenience we can choose respectively the {\bf R} scalar, {\bf Ric} and {\bf Riem} tensors (first $3$ terms in the above expression), without loss of generality. 

\subsubsection{Effect of the Gauss-Bonnet term in $D \ge 4$}

Another useful modification for us is given by substituting the last term with a generalized Gauss-Bonnet term in general dimension $D$, namely
\be
 	\mathrm{GB}_n \equiv R \Box^n R - 4 R_{\mu\nu} \Box^n R^{\mu\nu} + R_{\mu\nu\alpha\beta}\Box^n R^{\mu\nu\alpha\beta},
\ee
where the derivatives in form of box are centrally inserted between curvatures, and appear in the known decomposition of standard Gauss-Bonnet term. It happens that in $D=4$ the first variation of the Gauss-Bonnet term is a total derivative and hence the Gauss-Bonnet term does not influence the classical EOM. This justifies the Ricci-flat ansatz for the vacuum gravitational solutions in four dimensions in quadratic gravity theory with added Gauss-Bonnet term. It would be interesting to note what happens for the generalized GB, or in the other basis of operators quadratic in curvature. 

Similar situation happens if terms cubic and higher in curvature are taken into account. We can say something about the validity of this ansatz in the case of dimension four as well as in higher dimensions. It is easy to understand how the situation looks like here. In $D>4$ the original Gauss-Bonnet term contributes to the classical EOM and we find terms made only of the Riemann curvature tensor, hence Ricci-flat spacetimes generically will not be gravitational vacuum solutions of this theory. We see this on the level of EOM, but the same is true already in the action. In different basis with Weyl or Riemann tensor, the situation will be very similar, hence we expect non-Ricci-flat solutions in such higher dimensional quadratic in curvature theories. Our expectations are not changed, if we also include terms with third or higher powers of the curvature resulting from ordering the derivatives in initially quadratic gravitational curvatures actions. Typically there Ricci-flat manifolds will not be solutions. In $D = 4$ Ricci-flat manifolds will certainly be solutions of only terms quadratic in curvature actions and with no derivatives. Modification of theories beyond this simplest framework will generically not admit Ricci-flat solutions. 

It is interesting to notice that the similar conditions were found in the computation of tree-level amplitudes for the on-shell graviton scattering  \cite{scattering}. There the presence of terms with Ricci scalars or tensors (at least two powers on the level of Lagrangians) didn't give any impact on the result of computation in the Born approximation and the amplitudes were like in standard Einstein theory. This is in strong similarity to the situation with classical vacuum solutions. As we advocated above Ricci-flat spacetimes are solutions under the same conditions and these are the same solutions as in two-derivative Einstein gravity. However the explanation, which we found in \cite{scattering} was based on Anselmi's redefinition theorem. We understand that the conditions for on-shell gravitons in tree-level quantum gravity have much to do with the exact vacuum solutions of the original classical theory.

\subsubsection{Counterterms forced by a quantum theory}

Another case of interest is motivated by the quantum theory based on the Einstein-Hilbert action (E-H). It is known that this theory, if not coupled to matter, at quantum level is divergent at two loops and needed counterterms not present in the original action. This theory requires introduction of new counterterms, which were first discovered by Goroff and Sagnotti in 1982 \cite{GoroffSagnotti}. They contain terms that are cubic in curvatures and have the structure of {\bf{Riem}}$^3$ when various contractions of indices are used. Of course the classical theory based on such actions will presumably not admit Ricci-flat solutions, if there are no mysterious unexpected cancellations between terms on the level of EOM. This example is of great importance because it produces as first quadratic in curvature EOM with the explicit appearance of the Riemann tensor. Moreover if we require renormalizability in higher dimensions (starting from $D=6$) of quantum gravitational theory, then terms cubic in Riemann tensor must be there to absorb divergences at any loop order. Quantum gravity based on E-H action forces these terms to be present also in the two-loop effective action, hence it is of practical importance to find vacuum solutions in such theories because they will be the quantum corrected standard classical solutions known from E-H theory. 

Such corrections seem to be valuable for the signatures of quantum gravity in our world. It is strange to notice that precisely: Ricci-flat manifolds generally will not be solutions of such effective classical theory, which takes into account quantum effects based on quantization of standard Einstein-Hilbert action. Ricci-flat manifolds will be solutions only in the starting classical theory, but not in quantum theory based on it. This issue certainly deserves much more attention and deeper investigation.

\subsubsection{Killers}

One last comment is about killers (introduced after \eqref{K0})
which we typically use in the quantum theory to kill the beta functions
of the running gravitational coupling constants in one-loop exact super-renormalizable theories of gravity. 
It is always possible to choose these killers in such a way that they do not contain any Riemann tensor alone (and in any dimension).
It is crucial here to use at least two quartic killers (not cubic) and with at least two powers of Ricci curvatures \cite{modestoLeslaw}. This is always possible in any even dimension equal to or higher than $4$. Hence we conclude that Riemann tensor in such theories will not be present alone in the resulting EOM and that Ricci-flat manifolds are classical gravitational solutions also in these theories which are finite at the quantum level. There is of course a hidden assumption, that in the kinetic part of the action we use only operators quadratic in gravitational Ricci curvatures, but not contractions of Riemann, Weyl, Gauss-Bonnet etc there. To conclude we want to emphasize, that Riemann tensor in such theories will not be present alone in the resulting EOM.

\subsection{Delocalized matter}
 
The idea behind the usage of Ricci-flat solutions in general relativity was that they describe the spacetime where there is vanishing matter energy tensor. We know that if there is no gravitational field, then the flat spacetime is a solution. This is a unique solution from physical requirements of asymptotic flatness and standard topology of the universe. We also know that only the presence of matter can give rise to a non-trivial gravitational field, which is described by a non-trivial curved geometry with non-trivial Riemann tensor. Hence, we are led to the physical conclusion that Ricci-flat spacetimes work well as vacuum solutions only in some region of spacetime (like in the case of Schwarzschild solution). In other regions the matter must be present, otherwise the trivial flat spacetime would be the only physically admissible solution. Typically on the junction of two regions one specifies the boundary conditions saying how the metric tensor and its derivatives behave. For theories with higher derivatives we must specify conditions up to the order of the differential equation of motion. In the standard case (two derivative gravity) these Israel-Hawking conditions require continuity of the metric and its first derivative. If in the dynamics of the theory we have higher derivatives, then the metric must be in the class $\mathcal{C}^{N-1}$, where $N$ is the order of the differential equation. In a nonlocal theory the necessary conditions are up to infinite order, hence the metric understood as a real function must be smooth. This puts very tight constraints on the behaviour of the metric and actually one does not have any freedom in the process of sewing the solutions on both sides of the boundary. The solution inside (where the matter is present) uniquely determines the solution outside because of the smoothness condition. We can say in other words that matter is {\em delocalized} and its effect can not be restricted to a finite region of the spacetime.

Taking this seriously into account we conclude that, Ricci-flat manifolds, even if they respect all the symmetries of spacetime, are not admissible physical solutions for nonlocal gravitational theories. The usage of them in some regions of spacetime relies on the assumption that the impact of matter concentrated in some other regions can be encoded in finitely many numerical parameters, which must be matched on the boundary. Since matter is typically delocalized, this is not possible. If the matter is in some region of the spacetime, then in these theories the whole spacetime is not Ricci-flat. If there is no matter at all, then the flat spacetime solves EOM.

Another thing which requires commenting here is the issue of physical interest of these solutions in nonlocal theories (understood here as theories with infinitely many derivatives). 
In such theories we would have to specify infinite number of boundary (initial) conditions for our partial differential equations of the field theory. If we decide to specify them at one point, we would have to describe the behaviour of the solutions and all its derivatives (up to infinite order here) at this point (or we just specify its functional form in a neighborhood). This means that the boundary conditions are effectively nonlocal and this is precisely a weak type of nonlocality. Therefore, the following question can be asked: can we specify in a sensible way the boundary conditions on a surface that divides vacuum regions of spacetime from those where the matter is present? In nonlocal gravitational theories this is not possible, because typically matter undergoes {\em delocalization}. In other words in effective theory (based on gravitational Einstein equations) with some effective matter source everywhere in the spacetime 
\cite{ModestoMoffatNico, BambiMalaModesto, BambiMalaModesto2, calcagnimodesto, Yiwei, Cnl1}. If there is a seed with matter, then matter will get smeared out and will be present in every region of spacetime. 

More concretely, given any point-like source described by an energy tensor proportional to the Dirac delta distribution, it is delocalized by the nonlocal form factor in a Gaussian-like source, namely (for example)
\cite{BM, ModestoMoffatNico, BambiMalaModesto, BambiMalaModesto2, calcagnimodesto, Yiwei} 
\be
{\bf{\delta({x})}} \,\,\,\, \rightarrow \,\,\,\, {{e^{ - {\bf x}^2/l_{\Lambda}^2}}} \, , 
\ee 
where $l_{\Lambda}^2$ is the typical length scale of nonlocality.
This is the physical picture of what a nonlocal differential operator in action does on a localized (to some region) matter source. In this respect we are allowed to doubt whether our Ricci-flat solutions are sensible physical solutions to the EOM. 

An explicit calculation of the gravitational potential and some approximate solutions confirm the regularity 
of the spacetime in presence of matter \cite{ModestoMoffatNico, BiswasMazumdar, BM}. 
In short the EOM in the minimal super-renormalizable nonlocal theory reads as follows,
\be
{\bf G}  \sim  {\bf T} \,\,\,\,\,\, {\mbox{at large distances, and} } \,\,\,\,\,\, {\bf G}  \sim { e^{-H(-\Box_\Lambda)} \, {\bf T}} \sim 
{ \Lambda_{\rm eff} \, \bf g} \,\,\,\,\,\, {\mbox{at short distances} } ,
\ee
therefore, the exact solution will approach the Ricci-flat metric in the infrared regime, while in the high energy regime it approaches a de Sitter solution with effective cosmological constant 
$\Lambda_{\rm eff}$ depending on the original mass of the source, Newton constant $G_N$, and the nonlocality scale $\Lambda$. We got a de Sitter core in the short distances, because the effect of the delocalization at the center is to make the effective source with approximately constant and positive value of the energy density.

Moreover, we can exclude Ricci-flat solutions in nonlocal theories because typically they contain singularities. Of course these singular points do not belong to the spacetime, but still one can see the effects of the singularities when approaching them always being in the spacetime. 
Typically some curvature invariants tend to infinity during such approach. This is very undesirable feature of the solutions and it is unsatisfactory that such solutions exist at all in a nonlocal theory. 

We may think that nonlocal theories are on the best way to perform full desingularization of classical solutions known from Einstein theory. After all the main expectation is that in nonlocal theories, the {\bf local} singularities should disappear. However, Ricci-flat manifolds, which are still only mathematical solutions, develop such singularities and hence it is plausible that we find a physical motivation to exclude them. The exclusion of Ricci-flat manifolds does not occur in higher derivative gravitational theories and these are good physical vacuum solutions in the presence of matter in some or other regions of the spacetime, or in other words, the singularity is unavoidable in such theories.

In local higher derivative theories the distinction between regions
with matter or without matter can be kept sharp. In other words matter does not get delocalized in such theories and the argument of ``physicality and delocalization" exposed above does not apply. If we only look to the gravitational potential or other approximate specially symmetric solutions they can be regular also for local higher derivative theories. However, the delocalization argument exposed above here does not apply and we still have singular solutions.

\section{Maximally symmetric spacetimes}

In this section we will focus our attention on maximally symmetric spacetimes (MSS), which correspond to solutions with a cosmological constant, or in other words, to a uniformly distributed source of the energy momentum tensor that is proportional to the spacetime metric. These solutions are also known as Einstein manifolds because the Ricci tensor is proportional to the metric ${\bf Ric}=k {\bf g}$ and hence they describe vacuum solutions with the presence of cosmological constant. Later in the next section we will study solutions based on the FRW ansatz, which have obvious applications to cosmology. Both these classes of solutions can be viewed as vacuum solutions (being conformally flat) in Weyl gravity.
 
In this section we will be quite general and we will not specify the signature of the spacetimes; therefore, at the same time we would consider solutions in Minkowskian as well as Euclidean signature. 

\subsection{General criterion}

As it is well known in Euclidean signature we have two types of maximally symmetric spaces 
(hence uniformly curved), namely spheres and hyperboloids. They differ by the sign of their scalar curvature, which is respectively positive or negative. In the Minkowskian case we again have two types of maximally symmetric spaces, namely de Sitter and Anti de Sitter spaces. The sign of the curvature is as before. Since these solutions are non-trivial, because they contain matter, we will find them only in some special cases of generally covariant theory. One of the simplifying assumptions comes from the fact that on maximally symmetric spacetimes all curvatures are covariantly constant, because their tensor structures are expressed exclusively using the metric tensor. Of course we are here in metric theories, where any covariant derivative of the metric tensor vanish (no non-metricity). This means also that any covariant derivative acting on any curvature tensor on the manifold produces zero. This regards also covariant box operators. Thanks to this observation on the level of EOM we are left only with terms that do not contain any covariant derivatives. Hence the strategy here is to convert all possible curvature tensors to the commutation of derivatives (this was called ``uncommutation" in a previous section) because they would not influence the EOM, when the latter are evaluated on maximally symmetric spacetimes. In our EOM, as well as in the Lagrangians, we want the smallest possible number of curvatures and the highest number of derivatives. Let us consider things in order of increasing power of curvature appearing in the EOM, or on the level of Lagrangian. 

When we have only one curvature, we have the situation exactly like in Einstein-Hilbert theory. Then we know that to have MSS spacetimes, we must have a non-zero value of the cosmological constant parameter. Moreover, the precise equation of motion determines in this situation the sign of the curvature to be coincident with the sign of the cosmological constant. In more quantitative manner we describe the maximally symmetric spacetimes by giving their radius of curvature. In the case of the Einstein-Hilbert theory this parameter is inversely proportional to the value of the cosmological constant.

Our goal is to extend the preceding analysis for terms higher in curvature like those appearing naturally in the kinetic terms in our theory. Later we will comment on the effect of addition of higher than two power in curvatures. We start with terms quadratic in curvature on the level of the EOM originating from the action of higher derivative theory written explicitly in \eqref{appaction}. The degree of higher derivatives $n$ is kept fixed for the moment. To determine which types of terms are their origins on the level of Lagrangian we must notice one fact: when we take the (first) variation of the covariant derivative operator (or a Laplace type box operator), we never generate a gravitational curvature (see \eqref{var_scalar_box}, \eqref{var_tensorial_box}). Remembering that we do not see any effect of derivatives on curvatures in the equation of motion, we may restrict ourselves only to the consideration of the case with $n=0$ also on the Lagrangian level. This means, that we want to start with MSS solutions in quadratic gravity, which was first proposed by Stelle in 1977 \cite{Stelle}.

We want to describe these MSS by a parameter $A$ (see \eqref{MSS_EOM}) that is a proportionality factor between the Ricci tensor and the metric on MSS: 
\be
{\bf Ric} = A \, {\bf g}.
\ee
The parameter $A$ can be easily expressed in terms of the inverse square of the radius of the MSS and the dimensionality of the spacetime in question. Our goal here is to find an equation for the possible values of the parameter $A$ in terms of the parameters of the theory (like couplings, including the value of the cosmological constant). It will be shown that the equation relating the two things \eqref{trace2} will be of algebraic character. Due to the appearance of the parameter $A$ in every curvature tensor, we see that the exact EOM, which contains two powers of the curvature, will be quadratic in the $A$ variable. In more generality, if the EOM contain $n$ powers of curvature (any gravitational curvature), then this characteristic equation of MSS will be an $n$-th order algebraic equation for the unknown $A$. Let us analyze the structure of this equation in the case of quadratic gravity. We will see that much can be said even without the explicit derivation leading to it. Firstly we would like to notice that this equation is derived from parts of the first variation, which keeps all curvatures, so either from $\delta\sqrt{g}$, or from parts linear in curvatures from terms $\delta R$, $\delta R^{\mu\nu}$, but not from covariant Ricci tensor $\delta R_{\mu\nu}$. 
We can completely forget about varying covariant derivatives, or pieces, in which curvatures do not survive and are substituted by covariant derivatives on the fluctuations, as in \eqref{var_scalar_box} and \eqref{var_tensorial_box}. 

Note that it is not necessary to vary the covariant Ricci tensor, because its variation on the general background does not contain any background curvature \eqref{var_Ric}. We vary only the contravariant Ricci tensor, which amounts to varying of the contravariant metric tensors that were used to raise indices on this tensor. The resulting equations of motion are very simple in this case and we have written them below for the case $n=0$ \eqref{MSS_EOM}. The first two terms come from the Einstein-Hilbert theory with a cosmological constant. Our final equation for the characteristic \eqref{trace2} of MSS are obtained by taking the trace, or in other words by contracting with covariant metric tensor. In this way we will extract one significant scalar equation of motion. 

\subsection{Explicit form of the equation of motion}

After this brief analysis, let us take a closer look at the characteristic traced equation. 

An explicit form of the EOM, when the one form factor is present can help us in better understanding the content of the previous subsections. Taking the variation of the action with respect to metric tensor and making use of an implicit form for the
variation of the form factor we get the following result (compare appendix A.2),
\be \label{most_general_eom}
&& \hspace{-0.7cm}
E_{\mu\nu} 
 = \frac{ \delta \left[  \sqrt{|g|} \left( R -2 \Lambda_{\rm cc}+ G_{\alpha \beta} \frac{e^{H(-\Box_{\Lambda})} -1}{\Box} R^{\alpha \beta} \right) \right]}{\sqrt{|g|} \delta g^{\mu\nu}} \nonumber \\
 && = G_{\mu\nu} + \Lambda_{\rm cc} \, g_{\mu \nu} +  
 \frac{1}{\sqrt{|g|}} \frac{ \delta   \sqrt{|g|}  }{\delta g^{\mu\nu}} \left(G_{\alpha \beta} \frac{e^{H(-\Box_{\Lambda})} -1}{\Box} 
 R^{\alpha \beta} \right) 
 +   \frac{\delta G_{\alpha \beta}}{\delta g^{\mu \nu}  } \left( \frac{e^{H(-\Box_{\Lambda})} -1}{\Box} 
 R^{\alpha \beta} \right)  \nonumber \\
 && \hspace{1.1cm}
 +   G_{\alpha \beta} \frac{e^{H(-\Box_{\Lambda})} -1}{\Box}  \frac{\delta R^{\alpha \beta} }{\delta g^{\mu \nu}  }
 +  \frac{\delta \Box^R}{\delta g^{\mu\nu} }
 \left( 
  \frac{ \frac{e^{H(-\Box^G_{\Lambda})} - 1}{\Box^G} - \frac{e^{H(-\Box^R_{\Lambda})}-1}{\Box^R} }{\Box^R - \Box^G} 
 G_{\alpha \beta} R^{\alpha \beta} \right) \, ,
\ee
where $\Box^R$ acts only on $R^{\alpha\beta}$ and $\Box^G$ acts only on $G_{\alpha\beta}$. The operators $\Box^R$ and $\delta \Box^R$ do not commute and act in the indicated order \cite{Mirzabekian:1995ck}. It is easy to prove the following theorem for a form factor \eqref{Tombconstruction} without constant term in the IR expansion, i.e. $\omega_0=0$ (or equivalently $H(0)=0$),
\be
G_{\mu\nu} = -\Lambda_{\rm cc}  \, g_{\mu\nu} \,\,\,\, \Longrightarrow \,\,\,\, E_{\mu\nu} = 0 \,.
\ee
It is also trivial to prove the following claim,
\be
R_{\mu\nu} = 0, \quad \Lambda_{\rm cc} = 0 \,\,\,\, \Longrightarrow \,\,\,\, E_{\mu\nu} = 0 \,,
\ee
but the reverse is not true, as we have shown above in section 2. This theorem states that Ricci-flat manifolds are vacuum solutions of the theory (without cosmological constant). However, contrary to the situation in Einstein gravity, here they are not unique solutions. More about the compact form of the equation of motion can be found in the appendix.

We can make the general analysis of this section more quantitative using the EOM in the appendix. However the only case we are interested in here is when the form factor has a constant term in IR expansion (small $z$ expansion), or in other words when $H(0)\neq0$.
 For the de Sitter and Anti de Sitter spaces we have already said that 
 \be
 {\bf Ric} \propto {\bf g} \,\,\,\,\, {\mbox{and}} \,\,\,\,\, {\bf R} \propto {\rm const}.
 \ee
Therefore, all the terms involving the covariant derivative of either $R$ or $R_{\mu\nu}$ vanish on the left hand side of the EOM (\ref{YD-EOM}).
What is left is a polynomial in $R$ and $R_{\mu\nu}$
as evident from the explicit form of the tensors 
$\bf J$ and $\bf K$ that contribute to the EOM iff $n$, the power of box, vanishes. 
In this case the EOM reads as follows\footnote{We add to the action term $\omega_{\rm Riem}{\bf Riem}^2$, which is non-trivial in $D>4$ and which results from taking the constant term in expansion of form factor in ${\bf Riem}\,\gamma_4(\Box){\bf Riem}$.}, 
\bea \label{eq_12}
&&    \Lambda g^{\mu\nu} + G^{\mu\nu} + \delta^0_n \omega_{\rm R} \Big(\half g^{\mu\nu} R^2 - 2R^{\mu\nu} R \Big) 
\nonumber \\ 
&& + \delta^0_n \omega_{\rm Ric} \Big( \half g^{\mu\nu} R_{\alpha\beta} R^{\alpha\beta} - 2 R^{\beta \nu} R^{\mu }_{\ \ \beta} \Big) -2 \delta^0_n \omega_{\rm Riem} R^{\mu\alpha\beta\gamma} R^{\nu}{}_{\alpha\beta\gamma} = 0 \nonumber, \\
    \label{nzero}
\eea
that we can solve algebraically for $\omega_{\rm R}$, $\omega_{\rm Ric}$ and $\omega_{\rm Riem}$.
More specifically, consider the dS$_D$ and AdS$_D$ spacetimes whose embeddings are:
\be
    -x_0^2 + \sum_{i=1}^D x_i^2 = \alpha^2 \,\,\,\,\, {\mbox{for dS, and}} \,\,\, \,\,
    -x_0^2 - x_1^2 + \sum_{i=2}^D x_i^2 = -\alpha^2 \,\,\,\,\,\, {\mbox{for AdS}} \, . 
\ee
The Ricci tensor is $R_{\mu\nu} = (D-1)\alpha^{-2} \, g_{\mu\nu}$ (resp. $- (D-1)\alpha^{-2} \, g_{\mu\nu}$) for dS (resp. AdS). Defining $A \equiv (D-1)\alpha^{-2}$, hence $R = \pm DA$, $R_{\mu\nu} = \pm A g_{\mu\nu}$, $R_{\mu\nu\rho\sigma}=\pm \frac{2A}{D-1}g_{\mu [ \rho}g_{\sigma ] \nu}$ where the plus and minus signs refer to dS and AdS respectively.
Plugging the expressions for the {\bf R}, {\bf Ric} and {\bf Riem} into (\ref{nzero}) for the case $n=0$, we get
\bea
\hspace{-0.8cm}
&&    \Lambda g^{\mu\nu} \pm \Big( A g^{\mu\nu} - \half DA g^{\mu\nu} \Big) + \omega_{\rm R} \Big( \half g^{\mu\nu} D^2 A^2 - 2 D A^2 g^{\mu\nu} \Big) 
\nonumber
\\
&&+ \omega_{\rm Ric}\Big( \half g^{\mu\nu} D A^2 - 2 g^{\mu\nu} A^2 \Big) + \omega_{\rm Riem} \frac{A^2(D-4)}{D-1} g^{\mu\nu}= 0.
\nonumber 
\\
\hspace{-0.8cm} \label{MSS_EOM}
\eea
Taking the trace we get:
\be \bal
    0 &= \Lambda \pm \Big(1 - \frac{D}{2}\Big) A + \omega_R \Big( \half D^2 A^2 - 2 D A^2 \Big) + \omega_{\rm Ric} \Big( \frac{D}{2} A^2 - 2 A^2 \Big)  + \omega_{\rm Riem} \frac{A^2 D(D-4)}{D-1} \\
      &= \Lambda \pm \Big(1 - \frac{D}{2}\Big) A + \Big( \frac{D}{2} - 2 \Big) \Big(D \omega_{\rm R} + \omega_{\rm Ric} + \frac{2D}{D-1} \omega_{\rm Riem} \Big) A^2.
     \label{trace2}
\eal \ee

\subsection{The issue in general dimension}

We can make a posteriori observations for the first two terms originating from the theory without higher powers in curvatures. It is known, that the presence of the cosmological constant is inconsistent within Einstein-Hilbert theory in two dimensions. This implies that the contraction of Einstein tensor must  vanish in $2$ spacetime dimension, when evaluated on two-dimensional MSS. We know that in the definition of Einstein tensor, we use only up to one metric tensor, so in the contraction we may have at most one contraction between metric tensors (covariant with contravariant) resulting in one power of the spacetime dimensionality. Here instead of tracing we can extract covariant metric tensor, because equation of motion is proportional to it. We plug in the definition of Ricci scalar on MSS in terms of $A$ and $D$. Due to one instance of Ricci scalar in the definition of Einstein tensor, we see that this term can be at most linear in spacetime dimensionality. Equipped with this observation, we conclude that the coefficient in front of the contraction of Einstein tensor must be proportional to $D-2$. This is explicitly confirmed by direct computation, which shows that the proportionality factor is $-1/2$, because this was the coefficient in front of metric tensor in the definition of Einstein tensor.

Let's now consider the part of the characteristic equation quadratic in the unknown $A$. We see first that in all terms quadratic in curvature we have at most two powers of curvature scalars, which produce the square of the spacetime dimensionality. When Ricci tensor is rewritten in terms of metric, then another contraction of metric tensors is possible, but only one. This implies that this part of the equation is at most quadratic in the dimension of the spacetime $D$. Of course equation of motion are linearly proportional to the couplings of the theory appearing in the Lagrangian. We can say something more about the dependence on spacetime dimensionality by considering this time closer the case of four spacetime dimensions. Then we know that the Gauss-Bonnet term is a total derivative and does not influence equation of motion at all. This Gauss-Bonnet term can be written as a particular combination of terms quadratic in curvature. Let's choose to work in the simple basis with $R^2$, $R_{\mu\nu}^2$ and ${\rm Riem}^2$. Again the same argument says that, the coefficient in front of Gauss-Bonnet term must be proportional to $D-4$. Since we are working in different basis (with elements ${\bf R}^2$, ${\bf Ric}^2$ and ${\bf Riem}^2$), then the only way to get this proportionality for the coefficient for GB, irrespectively of the values of other coupling parameters ($\omega_R$ and $\omega_{\rm Ric}$, see the appendix), is that the effect of any term in the quadratic gravity action must be proportional to $D-4$. This is again confirmed by detailed computation for terms quadratic in $A$ of the characteristic equation for MSS. We indeed see the proportionality factor $D-4$. Another observation is that for four dimension the terms quadratic in $A$ completely vanish and the equation in quadratic gravity completely reduces to this one known in two-derivative Einstein-Hilbert theory. These all come from the fact that the Gauss-Bonnet term does not have any effect on the full equation of motion in four dimensions. Implications of this fact are profound. Again there is a linear relation between the inverse radius of curvature and the values of the cosmological constant in four dimensions. To have MSS as solutions there, we must have non-zero value of the cosmological constant parameter in the original action. This is the special property of quadratic gravity, MSS and four dimensions.

\subsection{Properties of the solution of the characteristic equation \eqref{trace2}}

Since in our convention parameter $A$ is always non-negative, this is the restriction must be put on the real solutions of the characteristic equation. Notice that this is a second order algebraic equation, hence about its solutions everything is known. First notice that $A=0$ is a solution only when $\Lambda=0$ where the theory is without cosmological constant. This is the flat spacetime solution, the conditions for which are analyzed in previous sections.

Let us write the quadratic equation for $A$ in the following form $aA^2+bA+c=0$. The coefficents $a$, $b$ and $c$ are expressed as functions of the parameters in the action (like $\omega_R$, $\omega_{\rm Ric}$ and $\omega_{\rm Riem}$) and the dimensionality of spacetime $D$. Everything below will be function of them. To find real solutions, we must require that the discriminant is non-negative,  $b^2-4ac \ge 0$. This condition is assumed to be valid from now on.

We shall facilitate our analysis by recalling the fact that the case of AdS differ from dS only by the flip of sign of $b$ term in the characteristic equation. This allows to collectively study both cases looking for positive $A$. We assume for simplicity of the analysis, that the parameter $a$ of the quadratic equation is positive. As shown above, in $D=4$ this parameter vanishes and we end up with linear instead of quadratic equation, exactly like in standard Einstein's theory. The analysis in the case $a<0$ will basically change the role of dS and AdS spacetimes. We notice that, due to the sign $\pm$ in front of $b$ coefficient and two solutions of the quadratic equation we have in principle always $4$ real roots of the characteristic equation. It happens that always out of them two are positive and two are negative. This is easily understood, because the absolute value of the $b$ parameter can be bigger than the square root (then the positive roots are different signs in solutions of quadratic equation) or can be smaller (then positive roots correspond to two possible signs of the $b$ parameter and always the plus sign for the root of quadratic equation). Actually in the first case we  choose this sign of $b$, which gives strictly positive value. Instead in the second considered case we use both signs of $b$. When $b=0$, the situation is a bit degenerate, because then we have only two roots, one positive and one negative, which are decided by the sign of the solution of quadratic equation. The conditions to have the situation first as described above is that $ac>0$, in such a way that the root of the discriminant is always smaller than the absolute value of the $b$ parameter. If $ac<0$, then we come back to the second case.

There is a physical meaning in these two distinct situations. In the first case sign of $b$ is fixed to be minus, hence $-b$ is always positive. In our theory the coefficient $b$ is always determined by Einstein-Hilbert theory, where there is only one contraction of curvatures, and that's why this ends up in a linear part of the equation, with the coefficient $b$. This coefficient is equal to $\mp\frac{D-2}{2}$. The origin of this factor was already explained before.
If we are above a critical dimension $D=2$, then this factor has fixed sign. The signs in front, however, correspond to dS and AdS spacetimes respectively. This means that we have two solutions for the radius of de Sitter spacetimes, while none existing solution for AdS. This is the situation for positive value of the coefficent $a$ in the equation. These two solutions for de Sitter differ in curvatures and the theory can not single out uniquely one solution. As we see in this case the other root typically corresponding to AdS spacetime became a less curved solution of the dS character. In the second case, when $ac<0$, we have solutions for different signs of the $b$ parameter and this tells us that we always find one solution for dS and one for AdS spacetimes. This is the physical reason, why to distinguish these two cases. Because the coefficient $c$ is always related to the value of cosmological constant (actually it is equal to) this divides set of MSS solutions of the theory depending on the sign of $\Lambda$. Roughly speaking for $a>0$ and $\Lambda>0$ we have solutions for both dS and AdS, while for negative cosmological constant we have only solutions for de Sitter spacetimes. This conclusion is true here, because remember that we are here in the case of quadratic gravity. Naively we could think, that it is totally opposite, to what is known in standard two-derivative gravitational theory, where for negative cosmological constant we have one unique AdS spacetime as a solution. 

We don' t want to neglect the case, when one of the roots is zero, $ac=0$. Partially this we have discussed above - flat spacetime solutions with vanishing value of the cosmological constant. But besides flat spacetime there is also another more interesting solution in this case. Typically in the case of quadratic gravity we will have to meet two distinct solutions with different curvature radii, two solutions will be admissible for a given set of coupling parameters and the value of cosmological constant. This is in distinction to standard case known from Einstein-Hilbert gravity, where for given set of parameters the solution was unique. In the case, when $\Lambda=0$ we have also another solution, which is necessarily dS spacetime. The radius of curvature of it is determined from a linear equation in this case. Again in the case of negative values of the coefficient $a$ the role of dS and AdS spacetimes is exchanged. As we see it was possible to say quite much about MSS solutions in quadratic gravity, which is actually a reduced theory for any theory with an action quadratic in curvatures. 

Now at the end we can speak a bit about generalizations of our analysis. From it start it was never designed to deal with higher in powers of curvature equation of motion (or Lagrangians). But the extension here is simple, we can neglect all terms with covariant derivatives etc. We concentrate only on terms with powers of curvature, then we vary them in such a way,  that in the resulting EOM we have only the same curvatures, without any derivatives. Substituting ansatze for MSS for all curvature tensors we derive algebraic equation for radius of curvature of these manifolds. For higher than quadratic powers of curvature this is complicated, so we will not present here analysis for the case of killer operators. Addition of any term with derivatives will not change our conclusion here, so we do not have to worry about the orders of derivatives or commutation of them.

For the MSS ansatz to be valid, it is required that the quadratic equation \eqref{trace2} in $A$ has at least one positive root.
In $D=4$, the quadratic term vanishes, and the situation is the same as in Einstein gravity. For $D \neq 4$, the roots of equation (\ref{trace2}) read
\be
    A = \dfrac{ \pm_1 \Big(\frac{D}{2} - 1\Big) \pm_2 \sqrt{\Big(1 - \frac{D}{2}\Big)^2 - 2\Lambda \Big(D - 4 \Big) \Big(D \omega_{\rm R} + \omega_{\rm Ric} + \frac{2D}{D-1} \omega_{\rm Riem} \Big) } }{ (D-4)\Big(D \omega_{\rm R} + \omega_{\rm Ric} + \frac{2D}{D-1} \omega_{\rm Riem} \Big) },
\ee
where $\pm_1$ accounts for equations of positive (dS) or negative (AdS) curvature, and $\pm_2$ indicates two possible solutions to each equation. Although exactly whether the solution is dS or AdS remains dependent on the specific values of coefficients, one can always conclude that if the discriminant is nonnegative, then at least one of the equations (dS and AdS) has a positive root, regardless of the sign of cosmological constant or whether it vanishes.

\section{Cosmological solutions}

As the last class of spacetimes we would like to discuss are the cosmological solutions in higher derivative theories 
(including nonlocal gravitational theories). The homogeneity and isotropy of space are assumed, hence it is possible to impose FRW ansatz for the metric in the comoving frame, namely 
\be
ds^2=-dt^2+a(t)^2\left(\frac{dr^2}{1-kr^2}+r^2d\Omega^2\right),
\ee
where $k=0,\pm1$ is the internal curvature in space. We are going to look for the reduced EOM satisfied by the scale factor $a(t)$ in theories with higher derivatives.

\subsection{The Weyl basis}

We first want to modify our theory a bit by going to another basis of terms, which decide the propagator around flat spacetime. Instead of using a basis with terms of the type $R\square^nR$ and $R_{\mu\nu}\square^n R^{\mu\nu}$, here we adopt a basis with terms $R\square^nR$ and $C_{\mu\nu\rho\sigma} \square^n C^{\mu\nu\rho\sigma}$
 ($n$ is a fixed positive or zero integer in a local theory, while we sum over it in a nonlocal theory). 
In $D=4$ a minimal Lagrangian of a ghost-free and super-renormalizable (or finite when suitable killer operators are added) theory of gravity reads as follow,
\be \label{weyl_basis_sec4}
&& \mathcal{L}_{\rm W} =  -2 \kappa_4^{-2} \sqrt{|g|}\Big[ R +
 C_{\mu \nu \rho \sigma} \gamma_{\rm C}(\Box) C^{\mu \nu \rho \sigma} + R \, \gamma_0(\Box) R \Big]  \,  ,
 \nonumber 
 \\
&& 
\gamma_{\rm C} = \frac{e^{H(-\Box_\Lambda)} -1}{2 \Box} \,\, , \,\,\,\, \gamma_{0} =  -\frac{1}{3} \gamma_{\rm C}  \, .
\ee
Theory written in this basis is {\emph not} equivalent to the theory written in the Ricci basis \eqref{appaction}. Their classical EOM and solutions are different, however their quantum properties are very similar and that is why we are motivated to consider the second theory here.

To find some non-trivial exact cosmological solutions in the presence of matter, it is convenient to change basis, and to use Weyl tensors instead of Ricci tensors or any others. The thing lies in the fact that the Weyl tensor (tensor of conformal curvature) vanishes on conformally flat manifolds. It happens that the cosmological metric given by the FRW ansatz is conformally flat (this is true in any number of dimensions). Therefore, the equation of motion resulting from such theories evaluated on the FRW ansatz contains only terms with Ricci scalars. In order to get them one needs to vary only the part with Ricci scalars, but not quadratic in Weyl tensors in the original action. The variation of the part with Weyl tensors will not have any impact on the equation of motion, if evaluated on the FRW ansatz. In the following we will concentrate on a very simple example to show how the procedure works. Later we will comment on possible generalizations of this setup by including more terms.

It could be interesting here to discuss some features of solutions in Weyl conformal gravity while we go to study theories written in Weyl basis.
We notice, in passing, that all vacuum solutions of standard Einstein gravity are also vacuum solutions of Weyl gravity in four spacetime dimensions described by the action $\int \!d^4x \sqrt{|g|} {\bf C}^2$. This is a simple consequence of the famous Gauss-Bonnet theorem on the level of action integral reducing the action to two terms quadratic in Ricci scalar and tensor respectively. In bigger generality trivial vacuum solutions in conformal gravity are conformally flat, because the theory admits conformal transformations of metric tensors as symmetries of the theory \cite{Mannheim}. All metrics in the same conformal class are seen as equivalent in such gravitational theory. The curvature is described by the tensor of conformal curvature, which is identified with the Weyl tensor. Gravitational vacuum in conformal gravity is defined in the same way as in standard gravity as a region in which there is no matter energy tensor. From Einstein EOM this is equivalent to the condition of Ricci-flatness. Deriving the EOM in conformal gravity it was discovered that the same role is played there by Bach tensor, which is defined as $B_{\mu\nu}=\left(\nabla^\rho\nabla^\sigma+1/2R^{\rho\sigma}\right)C_{\mu\rho\nu\sigma}$  \cite{solquadgrav}.

One can show by inspection of the above formula that Ricci-flatness of some region implies Bach-flatness, hence the original statement is also proven on the level of equation of motion. Therefore, we have the following types of vacuum solutions in conformal gravity: flat spacetimes, maximally symmetric spacetimes (conformally flat, but not Ricci-flat), Ricci-flat and the most non-trivial Bach-flat solutions. When in some region of spacetime a conformal matter is coupled to Weyl gravity, then the Bach tensor is non-zero in these regions and these latter ones constitute probably the most difficult solutions to obtain in conformal gravity. However, since Ricci-flat spacetimes are Bach-flat all the tests of gravity, in conditions where the matter is not very dense and does not exhibit big gradients and pressures, are passed by Weyl gravity too as well as did Einstein gravity. There is no a potential disagreement with the fact that Weyl gravity is a four-derivative theory, while popular solutions were found in Einstein theory with two derivatives. Simply in some special situations the other run-away solutions in Weyl theory can be forgotten and a subset of solutions coinciding with those from Einstein gravity can be used. Of course the full set of solutions is bigger in the theory with bigger number of derivatives. This last remark holds true also for the case of infinitely higher-derivative or weakly nonlocal theories.

Already here one should notice that the role played by the Weyl tensor for cosmological solutions is very similar to the role of Ricci tensor on Ricci-flat manifolds. FRW metrics are conformally flat in the similar sense like Ricci flat manifolds are. In both cases the gravitational field is present (and observers would experience tidal forces) because the Riemann tensor does not vanish. 
We think this is a very nice analogy between Weyl- and Ricci-flat conditions and their physical meanings. 

\subsection{A simple example}

Since we want to study non-trivial cosmological solutions with non-trivial matter source, the EOM will not trivially vanish when both sides of them are evaluated on the particular FRW ansatz. For this reason we will use some effective Lagrangians (see below) leading to the effective equation of motion that we hope to be able to solve exactly at the end. 

The solution can be further specified. As mentioned, in such situations the EOM will contain only expressions containing the Ricci scalar or the Ricci tensor and a bunch of covariant derivatives or covariant boxes \eqref{YD-J}. There is no Riemann or Weyl there; moreover, the number of Ricci tensors in each term of the expression can be only one -- they can originate only from the first variation of the Ricci scalar.
This is a distinction with the case of terms in the EOM resulting from terms of the type 
$R_{\mu\nu} \square^n R^{\mu\nu}$, where contractions of Ricci tensors are present. 
The form of the EOM is quadratic in the gravitational curvature; this observation will be important after taking trace, which will contract the indices of this only Ricci tensor, leaving us with a Ricci scalar.

Now we want to restrict the form of the matter in the cosmological setup. 
We will work exclusively with conformal matter, by which we mean matter whose trace of the energy tensor vanishes. The example in four dimensions 
of this kind of matter is electromagnetic radiation\footnote{Radiation as a conformal perfect fluid is defined in any dimension $D$ by barotropic index $w=\frac{1}{D-1}$. Electromagnetic radiation (vacuum electromagnetic field described by an action $\int\! d^Dx F^2$) is conformally-invariant only in $D=4$.}, which microscopically consists of massless (ultra-relativistic) particles. Hence here we will look for solutions with radiation, but without any other form of matter (nor dust, neither cosmological constant).

Another example of conformal matter can be found in inflation. When the potential of inflaton is $\lambda \varphi^4$, and we are in the slow-roll regime, the energy tensor of inflaton is trace-free, while the mass term for inflaton is not allowed because it would explicitly break conformal invariance. However, such potential for inflaton is not consistent with recent observations \cite{WMAP}.

For this conformal matter the full energy tensor does not vanish and hence the source side of the gravitational EOM does not vanish explicitly. However, for a conformal fluid (also like for any perfect fluid with provided equation of state) there exists a relation between the components of this tensor. They can be written in a diagonal form in the comoving frame where there is only energy density $\rho$ and pressure $p$. The first component is a purely timelike component of the energy tensor, while the second is the diagonal part of the spacelike sub-matrix, uniform in all spacelike components.

One modification that can be added here is the curvature of space that is encoded by the parameter $k$. The curvature of the space formally corresponds to a perfect fluid matter with the barotropic index $w=-\nicefrac{1}{3}$ in $D=4$. Therefore, we allow for exact classical solutions with non-trivial topology of the spatial slices of the universe (related to the curvature of spatial slices), but among the conventional matter source only radiation is taken into account. 

Due to the presence of the equation of state for the matter fluid we have only one parameter freedom on the matter side. Therefore, also on the gravitational side of the equation it is possible to restrict only to one representative equation. This can be taken as the trace of EOM \eqref{YD-J},
\be \label{traced_YD-J}
G^{\mu}_{\mu} + \omega_R \Big(- 6 \Box^{n+1} R - \sum_{i=0}^{n-1} \left( \nabla^{\mu}\Box^i R \right) \left( \nabla_{\mu} \Box^{n-1-i} R \right)
                                                    - 2 \sum_{i=0}^{n-1} \left( \Box^{i+1} R \right) \left( \Box^{n-1-i} R\right) \Big) = 0.
\ee
All the dynamical information is contained in one scalar equation for this particular metric ansatz and the form of matter fluid. After taking trace, we will get zero on the right hand side, while on the left side some expressions with Ricci scalar and derivatives acting on it are present. The possibility of having contraction of two Ricci tensors can be eliminated here, because there are not terms quadratic in them in EOM resulting from scalar part of the Lagrangian. Now this equation for Ricci scalars has very simple solution (but this may {\bf not} be the unique one), namely that Ricci scalar is zero. Then the trace equation will be satisfied automatically and trivially. On this solution for conformal matter one has not only vanishing Weyl tensor, but also vanishing Ricci scalar. The Ricci tensor $R^{\mu\nu}$ and Riemann tensors are non-vanishing, due to the presence of matter in our cosmological model. After all, they do not appear in the trace of \eqref{YD-J}, as long as we work in the Weyl basis \eqref{weyl_basis_sec4}.

Finding that vanishing curvature scalar is a solution here is not the end of the story, we should find some more detailed characteristic of the solution. To do this we can use the ansatz mentioned above that $R=0$. Plugging this into the full EOM it is found that all terms with higher derivatives vanish, because they are quadratic in curvature and there is always one power of scalar curvature, which is zero here. Therefore, only those originating from terms with one power of curvatures are left, or in other words the ones from Einstein-Hilbert action. It is true that in the final form of effective EOM we have only Einstein tensor and no higher derivatives when evaluated on this particular solution. It is obvious that effectively equation of motion reduces to Einstein equations. Also on the level of Lagrangian one can forget about higher derivatives for this solution.

For the conformal matter, some solutions which we were able to find are exactly the same like in standard two-derivative theory, by which we mean those for the scale factor $a(t)$ in all three cases for $k=1\,,0$ and $-1$ in Einstein-Hilbert gravity. Their dynamical behaviour is known and for example for flat space case there is a Big Bang singularity in the solution and similarly, the singularity cannot be avoided. Of course these are very special solutions and in principle one cannot exclude the possibility of some other viable solutions. However, here we are unable to find any reason to exclude these solutions, which are as physical as standard Friedmann solutions known in Einstein gravitational theory applied to cosmology, where matter (in a form of conformal matter) is present everywhere in the universe and the spacetimes in such solutions are not trivial. But we see that there is a singularity of the same type like in Friedmann solutions for radiation. There are also other possibilities for ansatze in cosmological framework discussed in \cite{koshe1}, which allow for analytic solutions.

\subsection{Possible generalizations}

Now we comment on possible generalizations of this simple setup.
The cosmological solutions mentioned above will also satisfy the EOM derived from actions containing higher than quadratic curvature invariants, the only requirement being that they must be constructed out of Ricci scalar or Weyl tensor exclusively. Actually on the level of EOM it is desirable to have at least one Weyl tensor in each term, which is bound to vanish on FRW ansatz. Therefore, on the level of Lagrangian one should deal with terms with at least two powers of Weyl tensors. The other tensors can multiply these, and covariant derivatives can act on such terms as well. These terms will not have any impact on the EOM, if they are put on particular background of FRW metric ansatz. The possibly non-vanishing terms contain only Ricci scalars or tensors for the success of our method.

It would be difficult to depart from conformal matter source because this would invalidate the $R = 0$ ansatz and hence would force us to solve full higher dimensional partial differential equation (even in one representative component). Therefore we think it is quite difficult to generalize our considerations away from the conformal matter case.
Regarding the remaining terms, on the level of EOM they must contain only one power of Ricci tensor, but arbitrary number of powers for Ricci scalar, while on the Lagrangian level we allow for more general terms (higher in curvatures), but the condition is that they still must contain {\em at most} one power of Ricci tensor. The Ricci tensor can be contracted with indices on covariant derivatives and can be further multiplied by other powers of the Ricci curvature scalar. Moreover in full generality, if we have conformal matter source then all conformally flat manifolds being solutions of Einstein's gravity (with the same matter content) are exact solutions of the theory \eqref{weyl_basis_sec4}. Once again, our analysis is not exhaustive and we cannot exclude possibilities that among terms forbidden (for the sake of simplicity) above some cancellations happen and equation of motion is satisfied. Only description of how to obtain some selected particular solutions is presented here.

Now we continue the discussion of the ambiguity initiated in \ref{ambiguity}. Of course the Riemann tensor in the EOM is not desirable at all, but it is not a problem, if its effect can be written entirely in terms of Weyl tensors and one Ricci tensor. Hence it is again not advised here to commute derivatives to produce new curvature tensors for the same reason as described in \ref{ambiguity} (typically by doing commutation a new Riemann tensor is produced and then if rewritten in terms of Weyl generates unwanted powers of Ricci tensors, which we always want to avoid here). As mentioned in \ref{ambiguity} our ansatz will not work for different order of derivatives, or if they originate from the use of Bianchi identities. Hence if one starts with the action written in a different basis, the classical theory will not admit the solutions found. We may consider the presence of generalized Gauss-Bonnet terms, but typically they will not be harmless to those found solutions. Only in $D=4$ will the original Gauss-Bonnet term preserve everything, because there it is a topological invariant.

The last issue concerns the usage of killers. It is possible to use them in such a form which will not spoil our solutions. 
They could tentatively look like $R^2\square^{n-2}R^2$ and $C^2\square^{n-2}C^2$, so differently from the standard postulated form in \eqref{K0}. One can show that they would do the killing properly as in the standard case. These new killers would be harmless to our just found cosmological solution, hence those will be also a solution in a theory finite on the quantum level.

The solution found here has a Big-Bang singularity that seems very difficult to avoid in the FRW universe in the presence of radiation. However, it is sufficient to have a small amount of dust matter and {\bf R} is no more identically zero, but actually again, like in the Ricci-flat case, we expect 
a delocalization of the energy density and likely a smearing of the initial singularity. Another very appealing possibility will probably be present when considering non-conformal perturbations. In such case the form factor will play again a crucial role. 

\subsection{Other solutions}

For a general higher derivative theory in the Weyl basis the action including the matter reads as follows,
\be \label{Weylbasisaction}
    S =  - 2 \kappa^{-2}_4 \int \!d^Dx \sqrt{|g|}  \Big(R - 2 \Lambda - \omega_{\rm R} R \Box^{n} R - \omega_{\rm C} C^{\mu\nu\rho\sigma} \Box^n C_{\mu\nu\rho\sigma} + \mathcal{L}_M \Big).
\ee
From the appendix the EOM for the FRW spacetime are \eqref{YD-EOM}:
\be \label{EoM}
    R^{\mu\nu} - \half g^{\mu\nu} R + \Lambda g^{\mu\nu} + \omega_{\rm R} J^{\mu\nu}
     = 8\pi G_N T^{\mu\nu},
\ee
where the matter energy momentum tensor is: $T_\nu^\mu={\rm diag}(\rho, p, \dots ,p)$. 

For the specific case of our nonlocal theory of gravity the discussion above simplifies as follows. Since the FRW spacetime is conformally flat, then the Weyl tensor is identically zero and we can focus on the EOM for the following reduced Lagrangian,
\be
\mathcal{L}_{\rm W} = - 2 \kappa_4^{-2} \sqrt{|g|}\left[ R
- \frac{1}{6} R \frac{e^{H(-\Box_\Lambda)} - 1}{\Box} R \right] .
\label{FRWtheory}
\ee
For radiation $T^\mu_\mu \equiv 0$ and tracing the equation of motion, it is evident from the EOM (\ref{YD-EOM}) that the trace of EOM is identically zero for the ansatz $R = 0$. If we now replace $R=0$ in the same equation of motion we turn out with the Friedman EOM for radiation. Therefore, the classical solution is still a valid solution for our nonlocal gravity.

In the theory \eqref{FRWtheory}, there also exist solutions with non-zero but constant Ricci scalar. First we notice that under the condition of constant $R$ and an absence of a constant term in the IR expansion of the form factor $\frac{e^{H(z)} - 1}{z}$, $J^{\mu\nu}$ in \eqref{YD-J} vanishes. To check the consistency of this ansatz, it suffices to take the trace of \eqref{EoM}, after which we see that only constant $T^{\mu}_{\mu}$ is consistent with constant $R$. In the trace EOM we do have terms quadratic in Ricci scalars, but only with derivatives, hence they vanish on the ansatz $R={\rm const}$. This restricts the matter content, and physically there are solutions only for situations where cosmological constant, radiation or curvature of space as matter content are present at the same time. Hence effective equation of motion looks exactly like this of Einstein gravity with cosmological constant.

For the convenience of the reader we present the form of these solutions here. If we define parameters $\alpha$ and $\beta$ such that $\Lambda = 3\beta^2 \neq 0$, $\alpha = \frac{10\pi}{3} G_N \rho_0$ with $\rho_0$ the density for radiation at the initial time $t=0$, the expressions for the scale factor and Ricci tensors are
\be \label{cosmo_sol}
    a(t) = \frac{ e^{-\beta t} } {2 \beta}  \sqrt{ \frac{1}{15} \left( e^{2\beta t}  + 15 k \right)^2 - 16\alpha\Lambda}    
\ee
\be \bal
    R_{00} &= -\frac{12\alpha}{5}a^{-4} - \Lambda \\
    R_{ij} &= \Big( -\frac{4\alpha}{5}a^{-4} + \Lambda \Big) g_{ij}\,,
\eal \ee
where the spatial metric equals to 
\be
g_{ij}=a^2(t){\rm diag}\left(\frac{1}{1-kr^2},r^2,r^2\sin^2\theta\right). 
\ee
From the traced equation of motion we read that
\be
    R = 4\Lambda\,.
\ee
Note that for $k = 1$, \eqref{cosmo_sol} can be well-behaving provided that $\alpha \Lambda < \frac{15}{16}$. In that case, $a(t)$ is everywhere nonzero, and the solution does not exhibit the Big Bang singularity; instead, we have a bounce at $t=0$.

It is also interesting to look at spacetimes not based on FRW ansatz. For the reader generally interested in nonlocal theories \eqref{FRWtheory} regardless of their good ultraviolet properties, we note 
that the G\"{o}del Universe \cite{Godel:1949ga}
\be
ds^2= \frac{1}{2\omega^2} [ -(dt + e^x dz)^2 + dx^2 + dy^2 + \tfrac{1}{2} e^{2x} dz^2]
\ee
 is an exact solution. Theories belonging to the class of \eqref{FRWtheory} are now understood not as an effective theory for FRW, but as our starting point, which are in general unitary but non-renormalizable, if considered as fundamental theories. Indeed for the G\"{o}del spacetime $R = {\rm const}$, but $R_{\mu\nu} \neq {\rm const}$, $J_{\mu\nu} = 0$ in the EOM. Moreover, for the Vaidya spacetime \cite{Vaidya:1951zz} it is true that $R=0$, so this is also an exact solution of the nonlocal theory (\ref{FRWtheory}). These metrics are solutions, if the form factor does not have a constant term in the expansion around $z=0$.

Note that the energy momentum sources for both solutions above are the same as those in standard Einstein-Hilbert gravity. They are solutions to theory \eqref{FRWtheory} provided that the constant term in the expansion of the form factor near $z=0$ is zero. This constant term has little to do with the UV asymptotic behaviour of the form factor, and the vanishing of this constant term is confirmed by the choice of Tomboulis form factor \eqref{Tomboulis} for any UV polynomial. Additionally we remark that the form factor as in equation \eqref{formfactor} has $\Box^{-1}$ term in the UV expansion, but not in IR.   The construction of the interpolating function by Tomboulis gives no constant term ($n=0$) in IR expansion even for the theory, which in UV tends to quadratic Stelle gravity and whose UV polynomial is $p(z)=z$. The reason for this is that the polynomial $p(z)$ does not have a constant term $p(0)=0$ and in the explicit construction \eqref{Tombconstruction} we use only square $p(z)^2$. 

\subsection{Conformal invariance}

We remark that the theory written in the Ricci basis possesses different cosmological solutions from the ones we found in the Weyl basis. It is obvious that on the classical level these two theories are not the same. However, on the quantum level they give rise to the same expression for graviton propagator and they differ slightly only in vertices resulting from the two theories. Therefore, the super-renormalizability and unitarity issues are solved in exactly the same way in these theories. For vacuum and MSS solutions we found it better to use the basis with Ricci tensors. Here however, we decided to change the theory and write it in the Weyl basis, because this is well suited for study of cosmological solutions.

From the previous discussion it is obvious that the presence of singularity in the cosmological context is inevitable. Nevertheless, we can propose the following resolution of this problem, which embodies the idea of conformal invariance of the quantum effective action. We believe that in full quantum theory the action will be conformally invariant, because only such actions can describe the theory at an ultraviolet (UV) fixed point (here it is assumed that the theory is UV complete or even finite). It is well-known that only conformal matter can be coupled to conformal gravity. On the other hand, FRW solutions are conformally flat metrics, so they do not require any matter source in a conformally invariant theory. If one still wants to describe cosmological models with a non-trivial source, the assumption of isotropy or homogeneity of space (i.e. the FRW ansatz) must be abandoned. This still leaves us with some freedom in choosing solutions suitable for cosmology, Lema\^{i}tre-Tolman-Bondi (LTB) metric being an example of spherically symmetric, but inhomogenous spacetimes. Another examples of homogenous, but anisotropic spacetimes are Bianchi models. This accounts for the fact that the FRW solution just found is not a very physical one.

Since FRW spacetime with whatever form of the scale factor $a(t)$ (in particular possessing Big Bang singularity) is equivalent to the flat spacetime in conformal gravity, the problem of singularity disappears, because obviously on flat spacetime singularities do not occur. This idea of resolving the singularities by conformal transformations originated from Penrose's ideas \cite{Penrose} \footnote{In solutions of the form \eqref{cosmo_sol} found earlier, there are two different regimes. At large $t$, the solutions are cosmological constant-dominated. When $t$ is close to zero, there is Big Bang singularity. However, we notice that in the early stage of the evolution the solutions are dominated by conformal matter. Therefore, it is possible to remove such singularity using the same arguments (related to conformal transformations) in conformal gravity. Note, however, that such spacetimes are only approximate solutions to the equation of motion derived from the action principle of conformal gravity.}.

We want to comment on the FRW solutions in the classical action of quantum gravity as proposed in \cite{AnselmiWeylaction}. The theory \eqref{Weylbasisaction} may be generalized even further by including terms higher in Weyl curvature (than quadratic) and containing more covariant derivatives. In \cite{AnselmiWeylaction} similar theory was considered as a basis for classical action of quantum gravity. There are no terms higher in Ricci scalar than the Einstein-Hilbert term and all the other gravitational curvatures are in Weyl tensors. It was found that all FRW spacetimes are solutions to the theory in \cite{AnselmiWeylaction}. This finding can be generalized to any (locally) conformally flat spacetime, namely for the theory all conformally flat solutions of Einstein's theory are also solutions there with the same matter content. All these properties come from the fact, that on conformally flat manifolds Weyl tensor vanishes. However, in the theory \eqref{Weylbasisaction} this last statement is not generally true as we have shown above. The reason is the presence of terms of the type $R\Box^n R$. If we have conformal matter source then all conformally flat manifolds being solutions of Einstein's gravity (with the same matter content) are exact solutions of the theory \eqref{Weylbasisaction} with generalizations first introduced by Anselmi in \cite{AnselmiWeylaction}.

We feel obliged to comment on the following issue. In the work of Anselmi \cite{AnselmiWeylaction} it was shown, that the action he proposed is the most general classical action of quantum gravity up to the field redefinitions defined using Einstein's equations of motion. In such theory conformally flat spacetimes are solutions for some generally non-vanishing matter source. In conformal gravity they are vacuum solutions of the theory, so matter energy tensor is zero everywhere. This is only in apparent contradiction to the similar statements in conformal gravity, which is also candidate theory emboding quantum effective action of gravity. In tentative form of conformal quantum gravity we expect terms as in the action \eqref{Weylbasisaction}, which are quadratic in curvatures, supplemented by other terms higher in curvature, which indeed may be constructed using only conformal curvature tensors. This is the difference between this action of conformal gravity and the one proposed by Anselmi. According to \cite{AnselmiWeylaction} terms quadratic in curvature can be redefined on-shell, this is genuinely true using Einstein's equations of motion. Since in conformal gravity we have different EOM and since Einstein gravity is not conformally covariant in $D=4$, then by applying field redefinition method we lose conformal symmetry in the theory. This is already obvious in the simple example of Weyl gravity, described by the Lagrangian ${\bf C}^2$, where field redefinition to E-H gravity is possible, but the final theory does not enjoy conformal symmetry anymore. Therefore, all our statements about resolutions of cosmological singularity problem in conformal quantum gravity hold true and do not disagree with conclusions of Anselmi.

\section{Multi-scale black holes in vacuum}
We explicitly showed that Schwarzschild or Kerr black holes are exact solutions of a large 
class of local higher derivative theories and, in particular of our candidate super-renormalizable theory.
We also stressed that probably such black hole solutions are not physical in nonlocal theories because 
the delocalization is here at work. However, mathematically they are solutions and we can speculate 
about some astrophysical implications.
We also note the possible existence of black hole solutions of different sizes and same mass, due to new energy scales present in the higher derivative theory.
We remind that in Einstein gravity the Schwarzschild metric obtained by integrating the EOM, ${\bf Ric} =0$, shows a free integration constant that is fixed to be the Schwarzschild radius $r_s = 2 G_N M$ 
in order to read out the gravitational Newtonian potential from the $g_{00}$ component of the metric.
The Einstein equations are second order and we only have the scale of the Newton's constant $G_N$ in the game, so we only have the reminded above Schwarzschild radius of the event horizon. 
As already pointed out in this theory we can have Schwarzschild (or Kerr) black holes with standard locations of horizon, whose scales are characterized by the dimensionful $G_N$.

Of course, we have $G_N$ in our theory too and so the usual black holes of Einstein theory are present there, provided the conditions from subsection \ref{critRicflat} hold true. However, in higher derivative theories
there are new mass scales $\Lambda_i$ ($i = 1, \ldots, k$, with $k$ related to the higher derivative order of the theory)
and in principle we can not exclude the following Schwarzschild-like {\em vacuum solutions} 
\be
ds^2 = - \left( 1 - \frac{r_{\Lambda_i}}{r} \right) dt^2 + \frac{dr^2}{\left( 1 - \frac{r_{\Lambda_i}}{r} \right)}
+ r^2 d \Omega^{2} \,\, , \,\,\,\,\,\,\,\,\, r_{\Lambda_i} = \frac{2M}{\Lambda_i^2} \,.
\ee
In addition to spherically symmetric black holes with horizons given by $r_{\Lambda_i}$ we can also consider Kerr-like black holes characterized by different scales than $G_N$. It is crucial to consider vacuum spacetime solutions whether we want to support the idea of gravitational multi-scales characterizing these solutions. By multi-scale solutions we mean here these one with gravitational scales different from the standard one given by $G_N$. Only in the case of higher derivative gravity without extra fields permeating the spacetime, a purely mathematical vacuum solution leaves us free to fix the extra scale as an extra gravitational radius. However, it remains to be seen whether these solutions are viable from the astrophysical point of view.

Assuming at the moment physical viability of the above multi-scale black hole solutions, we can evaluate the Hawking temperature of Schwarzschild-like black hole characterized by a new gravitational scale $G_{\Lambda_i}={\Lambda_i}^{-2}$, namely 
\be
T_{\Lambda_i} = \frac{1}{8 \pi G_{\Lambda_i} M} = \frac{\Lambda_i^2}{8 \pi M}\,, \,\,\,\,\, \mbox{ while in the standard theory } \,\,\,\,\, T_H = \frac{1}{8 \pi G_N M} = \frac{M_p^2}{8 \pi M} \,.
\ee
The black hole lifetime will be shorter or longer depending on the value of the scales in the problem. For example for the Starobinsky theory $R+ R^2/6 m^2$ \cite{Starobinsky}, the new scale is actually smaller than the Planck scale $M_p$, namely $m \sim 10^{-5} M_p$. If  $\Lambda_i < M_p$, the black hole's Hawking temperature is lower and its lifetime is longer than in the standard case. Therefore, if they were created in the early universe, through a mechanism not easily identifiable, 
they will evaporate later in time. Here we assume the Starobinsky model to be the best candidate to explain early universe inflation. Therefore, 
if a mechanism exists to create such compact objects in the epoch without any matter, then we should expect in the universe a population of black holes with the same mass, but event horizon $10^5$ times bigger and Hawking temperature $10^{-10}$ times lower than the standard one.

In a nonlocal theory multi-scale black holes cannot be created by gravitational collapse because it is the Newton constant $G_N$ that appears as an overall coupling in the gravity sector, and finally in coupling gravity with matter,
namely,
\be
\frac{1}{8 \pi G_N} E_{\mu\nu} = T_{\mu\nu}\, , \,\,\, {\rm where \ } E_{\mu\nu} \mbox{ is as indicated in \eqref{most_general_eom}.}
\ee
The new mass scale $\Lambda$, appearing in \eqref{formfactor}, is responsible for the delocalization of source and in the result exact Ricci-flat black holes with different than $G_N$ gravitational scales are not physical solutions. On the other hand, for higher derivative theories we do not have delocalization and we do not see any reason why to exclude singular spherically symmetric spacetimes, possibly with various gravitational scales\footnote{
Of course, in higher derivative theories we can also
have regular solutions as shown in the calculation of the gravitational potential \cite{Tiberio}.}.

\section{Conclusions}

In this paper we considered exact solutions for 
the classical equation of motion of local and nonlocal higher derivative gravitational theories. Greater attention has been devoted to the following two classes of nonlocal theories, which both enjoy the same properties at quantum level: unitarity and super-renormalizability or finiteness, 
\be
&& \hspace{-1cm} 
\boxed{
\mathcal{L}_{\rm g} = -  2 \kappa_{D}^{-2} \, \sqrt{|g|} 
\Big[ \Lambda_{\rm cc} + {\bf R} 
+
{\bf R} \, 
 \gamma_0(\Box)
 {\bf R }
 + {\bf Ric} \, 
\gamma_2(\Box)
 {\bf Ric} 
+ {\bf {V}} 
\Big]\,, }
   \label{RicciBasis} \\
&& \hspace{-1cm} 
\boxed{ \mathcal{L}_{\rm W} =  -2 \kappa_4^{-2} \sqrt{|g|}\Big[ {\bf R} +
 {\bf C} { \gamma_{\rm C}(\Box ) } {\bf C} + {\bf R} \,   \gamma_{0}(\Box)  {\bf R} 
 + {\bf V(C)}
 \Big]\,,  }
 \label{WeylBasis} 
 \\
&& \hspace{-.8cm} \mbox{with} \;
\gamma_{\rm C} = \frac{e^ { H\left( - \Box_\Lambda \right)}  -1}{2 \Box} \,\, , \,\,\,\, \gamma_{0} = - \frac{1}{3} \gamma_{\rm C}  \, ,  
\ee
where $\gamma_2$, $\gamma_{\rm C}$ and $\gamma_0$ are entire functions (form factors) of the covariant d'Alembertian operator.

Let us start summarizing the results about the first class of theories (\ref{RicciBasis}). Specifically, it has been shown that a large class of exact solutions of Einstein gravity contains also solutions for the theory with higher derivative terms. By virtue of our construction, when only the Ricci tensor and Ricci scalar appear in the classical action, the EOM always consists of monomials of the Ricci curvature tensor and derivatives thereof. Hence it was quite obvious that Minkowski and Ricci-flat spacetimes are classical solutions in a region without matter content. Therefore, for example the most famous and well tested Schwarzschild spacetime is an exact solution of the nonlocal theory (\ref{RicciBasis}). Moreover, Kerr  spacetime and all other vacuum solutions of Einstein gravity are also exact solutions here. If the Riemann tensor was present in the construction of a Lagrangian, then the Ricci-flat ansatz might fail unless under special circumstances, for instance when $\mathbf{Riem}$ appears in the particular combination of the Gauss-Bonnet term. We also commented on the impact of Goroff-Sagnotti-like terms and killers of the beta functions on the Ricci-flat solutions in the theory.

Despite the agreement with Einstein gravity on Ricci-flat spacetimes, these higher derivative theories have some novel features. It is known that curvature of spacetime is determined by matter content and distribution, and therefore the Ricci-flat ansatz can describe only part of the spacetime. In fact the simplest vacuum solutions usually possess singularities. Regarding this problem, a~nonlocal higher derivative theory might come as a rescue because of the presence of a weakly nonlocal differential operator. This operator, which effectively delocalizes the matter content and smears out the singularities to some smooth core, is most often constructed based on the covariant box operator.

Next class of solutions of interest are maximally symmetric spacetimes where the Ricci tensor is proportional to the metric with a constant coefficient, ${\bf Ric} \sim {\bf g}$. The examples in Lorentzian signature are de Sitter and Anti-de Sitter spacetimes. In this case the EOM greatly simplifies because the derivatives of the Ricci tensor vanish. The resultant equation of motion is at most a quadratic algebraic equation from which the value of the Ricci scalar and hence the characteristic radius of curvature can be obtained in terms of the cosmological constant, the spacetime dimension, and the coefficients of the higher derivative terms. The de Sitter or Anti-de Sitter solutions are generally exact solutions in any spacetime dimension $D$. We notice that a bit counterintuitively under some special conditions and in $D>4$ (A)dS spacetimes are solutions with (positive) negative value of the cosmological constant. For the case of nonlocal form factors with $H(0)=0$ or in $D=4$ the characteristic equation reduces to the linear one identical to the one in Einstein gravity.

For the case of cosmological models we decided to work in the Weyl basis (\ref{WeylBasis}), where the conformal flatness of FRW spacetimes is exploited. We found that the presence of conformal matter as a source is consistent with the same solutions as in Einstein gravity and the Big Bang singularity seems to be unavoidable. The simple modifications are here: inclusion of curvature of space and addition of cosmological constant fluid to the matter content of the model, however the conclusions remain unchanged.

Other interesting cosmological models like the G\"{o}del universe or the Vaidya metric for the gravitational collapse also fit into the general framework of weakly nonlocal gravitational theories. In particular they are solutions of an incomplete theory that can be obtained from the Lagrangian (\ref{WeylBasis}) by removing (by hand) the term quadratic in the Weyl tensors
${\bf C} \gamma_{\rm C} (\Box) {\bf C}$. 

While FRW models are physically meaningful solutions for describing the universe, they are rather trivial (conformally flat) from the point of view of conformal gravity that is a possible approach to solve the singularity problems brought up by Penrose. Coupling to non-trivial conformal matter would spoil the isotropy and homogeneity of the space, rendering the FRW solutions uninteresting. However, the validity of conformal gravity for describing the universe still is not fully accepted. 

Due to one or more new independent mass scales introduced in the higher derivative action, it was speculatively suggested that new black holes characterized by the new mass scales may populate the universe. We can for instance read off this mass scale from the Starobinsky model for inflation in order to be consistent with the cosmology of the very early universe. In principle we have black hole solutions characterized by this lower mass scale.  These black holes cannot, however, result from gravitational collapse of any matter. As a consequence of new intrinsic gravitational scale such black holes are with much larger radius of the event horizon and much lower Hawking temperature. This is the main reason why to consider this interesting idea of multi-scale solutions, consistent from the mathematical point of view, but very speculative from the physical point of view. 

In this work we have found solutions in higher derivative and nonlocal gravity theories only identical to solutions in two-derivative gravity. However we know that besides them in the theory exist for sure other exact solutions not shared with Einstein theory. Among many we can mention here: Starobinsky inflationary solutions \cite{Starobinsky}, solutions with delocalized matter, and solutions in Weyl \cite{Mannheim} and Stelle quadratic gravity \cite{solquadgrav}. Only in the cases of globally empty spacetime and filled unformly by cosmological constant we are sure that unique solutions are flat and maximally symmetric spacetimes respectively. In a general class of theories considered in this paper vacuum solutions may not be necessarily Ricci-flat and similarly cosmological spacetimes with radiation as a matter may not be consistent with condition of vanishing of the Ricci scalar. We remark that these last two conditions were just useful ansatze used in this paper to find some solutions.

Finally we want to summarize the issues of the presence of singular solutions in these theories. It is commonly argued that nonlocal theories are in general able to tame spacetime singularities, nevertheless here we have a large class of  exact solutions with singularities. We think that our results can only be taken seriously for distances larger than the characteristic scale of nonlocality $\Lambda^{-1}$. At very short distances the solutions are not physically viable because a ``delocalization mechanism" is at work. The form factor acts on the matter sources turning them into smooth distributions with bigger support spread everywhere in the spacetime. More concretely, any point-like source, described by an energy tensor proportional to the Dirac delta $\bf{\delta({x})}$, usually giving rise to singularities in Einstein gravity, is delocalized into a Gaussian-like effective source \cite{BM, ModestoMoffatNico, BambiMalaModesto, BambiMalaModesto2, calcagnimodesto, Yiwei}, for example
\be
 \quad e^{\Box_\Lambda} {\bf{\delta({x})}} = 
 {e^{ -{\Lambda}^2 {\bf x}^2}} \, . 
\ee
For classical localized $\delta({\bf x})$-like sources, the special feature of physical solutions of the Lagrangian (\ref{RicciBasis}) is to interpolate between the Minkowski vacuum at large distances and the de Sitter core asymptotically at short distances. In conclusion in these cases the singularity is resolved.

We consider the exact solutions found in this paper very interesting from the mathematical point of view and also for the physics in the infrared regime, but we expect a very different and non-singular structure of the spacetime will be manifest in the UV regime. Moreover, a general feature of the theories in (\ref{RicciBasis}) is asymptotic freedom, which is in agreement with effective flatness at distances much shorter than the de Sitter radius. For the case of FRW homogeneous and isotropic spacetimes, asymptotic freedom together with a soft propagator in the UV imply a bouncing (singularity-free) universe in the case of cosmology and/or in the gravitational collapse case \cite{Frolov, Tiberio, BambiMalaModesto, BambiMalaModesto2}. However, those last FRW solutions are just approximate solutions. Here we are considering the theory written in the Ricci basis, i.e. \eqref{RicciBasis}. For a conformally invariant theory to be discussed below, the mechanism to solve the singularity problem is different and based on the conformal invariance itself.  

We can ask the the following question: what about exact cosmological solutions based on the FRW ansatz? Are they regular or singular in theories \eqref{RicciBasis}? There is an extensive literature about exact solutions in nonlocal theories \cite{BiswasMazumdar, koshe1}, but here we are interested in looking for singular spacetimes in theories well defined at quantum level.
Indeed, as reminded above, we would like to confirm or disprove the following conjecture:

``{\em finite theories at quantum level are singularity-free at classical level} ". 

For this goal we consider the theory (\ref{WeylBasis}) in the Weyl basis, which is here proposed for the first time.
This theory is ghost-free and super-renormalizable or finite at quantum level depending on the front coefficients for the killer operators (not undergoing renormalization at quantum level) in the potential ${\bf V( C ) }$. We note right at the beginning that Ricci-flat spacetimes are not exact vacuum solutions here, contrary to the case of theories written in the Ricci basis. Therefore spacetimes with Schwarzschild-like singularities are not problematic, because these solutions can be viewed at most as approximate valid only at large distances. 

For the FRW spacetimes the Weyl tensor is identically zero and when conformal matter (radiation) is coupled to the gravitational theory (\ref{WeylBasis}) only the term quadratic in the Ricci scalar and the Einstein-Hilbert term give contribution to the EOM. However, for conformal matter the trace of the energy tensor $\bf T$ is identically zero and taking the trace of the EOM
for the reduced theory (\ref{FRWtheory}) we also find ${\bf R} =0$ as a consistent ansatz. We end up with exactly the same reduced EOM as for 
Einstein gravity and the usual Big Bang singular solution for a radiation dominated universe seems unavoidable. For the universe filled with radiation source is distributed everywhere uniformly in space, hence delocalization cannot help here in removing the initial singularity.

At this level the conjecture seems disproved because here we can not use the delocalization arguments invoked above for localized sources. However, here we must be careful before reaching hasty conclusions. If the theory is only super-renormalizable we still have singularities at quantum level (loop divergences) and the above conjecture is not disproved. On the other hand the finite theory is scale-invariant at quantum level (without divergences) and we expect the quantum action to enjoy full conformal symmetry when all the finite contributions are taken into account\footnote{It was proven in \cite{Polchinski} that any reasonable unitary 4-dimensional quantum field theory enjoying scale-invariance can be easily promoted to a conformal field theory.}. Therefore, the chain of implications, just mentioned above for FRW spacetimes, 
\be
{\rm tr\, {\bf{T}} } = 0 \ \Longrightarrow \ {\rm consistent\ ansatz\ } {\bf R} = 0 \  \Longrightarrow \ {\bf G} = 8 \pi \, {\bf T}  
\ \Longrightarrow \ \mbox{Big Bang Singularity} 
\ee
does not apply anymore. This happens because despite that reduced EOM is not conformally covariant, the full theory is invariant and hence conformal transformations are allowed as symmetry transformations of the theory. They map solutions found from reduced EOM (coinciding with Einstein equation) to other field configurations, which are not solutions of the same reduced EOM (with the same matter content), but which are good solutions in the full theory.
Actually, in the conformally invariant theory all the FRW spacetimes belong to the same equivalence class of conformally flat spacetimes, hence singular and regular ones are equivalent. In this theory with a conformal transformation we can always map a singular solution into a regular one. 

Even though we have full knowledge of the scale-invariant theory, the conformally invariant extension is at the moment under careful investigation at quantum  as well as at classical level\footnote{Here the conformally invariant quantum effective action is viewed as a classical one, hence it makes sense to perform classical conformal transformations to its solutions.}. If we succeed, the outcome will be an explicit realization of the Penrose conjecture for cyclic conformally invariant cosmology \cite{Penrose}. 

 \appendix \section{Appendix. Equation of motion}

\subsection{EOM in higher derivative gravity}
We here evaluate the EOM for a general higher derivative gravitational theory in a $D$-dimensional spacetime. The result can be directly applied to the weakly nonlocal theory in the Ricci basis, but not to the theory written in the Weyl basis. The action has the following form, 
\be
\label{appaction}
S = - 2 \kappa^{-2}_D \int\! d^Dx \sqrt{|g|} \Big(R - 2 \Lambda_{\rm cc} - \omega_{\rm R} R\, \Box^{n} R - \omega_{\rm Ric} R^{\mu\nu} \Box^n R_{\mu\nu} \Big),
\ee
where $\Lambda_{\rm cc}$ is the cosmological constant, $R_{\mu\nu}$ is the Ricci curvature tensor, and $R$ denotes the Ricci scalar. In formula above $\omega_{\rm R}$ and $\omega_{\rm Ric}$ are dimensionful parameters measuring the strength of $O({\cal R}^2)$ terms. Let $h_{\mu \nu} = \delta g_{\mu\nu}$ denote the variation of the covariant spacetime metric. In terms of $h_{\mu \nu}$, the change in the action $S$ is
\be
 \delta S = \int\! d^Dx \sqrt{|g|} E^{\mu\nu} h_{\mu\nu},
\ee
where ${\bf E}$ is a rank-2 symmetric tensor, we want to calculate. The EOM will simply amount to $E^{\mu\nu} = 0$.

To make easy the computation several definitions and simple identities are listed below for the reader's convenience:
\be 
\label{building-block}
&&    G^{\mu\nu} = R^{\mu\nu} - \half g^{\mu\nu} R \, , \,\,\,\,  \\
&&    \delta \Gamma^\rho_{\mu\nu} = \half g^{\rho\sigma} \Big( \nabla_\mu h_{\sigma \nu} + \nabla_\nu h_{\mu \sigma} - \nabla_\sigma h_{\mu\nu} \Big), \\
    && \delta \sqrt{|g|} = \half \sqrt{|g|} g^{\mu \nu} h_{\mu \nu}, \\
    && \delta g^{\rho\sigma} = -g^{\mu\rho}g^{\sigma\nu}h_{\mu\nu}, \\
   \label{var_Ric} &&  \delta R_{\mu \nu} = \nabla_\rho \delta\Gamma^\rho_{\mu\nu} - \nabla_\nu \delta\Gamma^\rho_{\mu\rho}
                       = -\half \Box h_{\mu \nu} + \nabla^\sigma\nabla_{(\mu} h_{\nu) \sigma } - \half g^{\rho\sigma}\nabla_\nu\nabla_\mu h_{\rho\sigma}, \\
    &&  \delta R^{\mu \nu} = \delta \(g^{\mu\rho}g^{\nu\sigma} R_{\rho\sigma}\)= -2 h^{\rho(\mu} R^{\nu)}_\rho +g^{\mu\rho}g^{\nu\sigma}\delta R_{\rho\sigma} \nonumber \\
 \label{var_Riccontr} && \hspace{0.85cm} = -2 h^{\rho(\mu} R^{\nu)}_\rho-\half \Box h^{\mu \nu} + \nabla^\sigma\nabla^{(\mu} h^{\nu)}_{ \sigma } - \half g^{\rho\sigma}\nabla^\nu\nabla^\mu h_{\rho\sigma},\\ 
   \label{var_Rs} && \delta R = \delta (g^{\mu\nu} R_{\mu\nu})
             = -R^{\mu\nu} h_{\mu\nu} - g^{\mu\nu}\Box h_{\mu\nu} + \nabla^\mu\nabla^\nu h_{\mu\nu}, 
   \label{var_scalar_box} 
   \ee
   \be
   &&
    (\delta \Box) \phi = \delta (g^{\mu\nu} \nabla_\mu \nabla_\nu \phi)
                    = - \(\nabla^\nu h_{\mu\nu}\)\nabla^\mu \phi + \half  g^{\mu\nu} \(\nabla_\sigma h_{\mu\nu}\) \nabla^\sigma \phi - h_{\mu\nu}\nabla^\mu\nabla^\nu \phi,
 \ee
where the $\phi$ in the last equation is some general scalar defined on the spacetime manifold. It will take the form of $\Box^n R$ in following calculations. Moreover we need also the first variation of the covariant box operator acting on a general symmetric tensor with two covariant indices $\alpha_{\mu\nu}$. This is given by
\be
    \label{var_tensorial_box}
    && (\delta \Box) \alpha_{\mu\nu} = -h_{\rho \sigma} \nabla^\rho \nabla^\sigma \alpha_{\mu\nu} - \half \Bigg( \nabla^\eta \Big( \big( \nabla_\eta h^\lambda_\mu + \nabla_\mu h_\eta^\lambda - \nabla^\lambda h_{\eta \mu} \big) \alpha_{\lambda \nu} \Big) \nonumber \\
                                        &&\hspace{1.7cm} 
                                        + \nabla^\eta \Big( \big( \nabla_\eta h^\lambda_\nu + \nabla_\nu h_\eta^\lambda - \nabla^\lambda h_{\eta \nu} \big) \alpha_{\lambda \mu} \Big) 
                                        + \Big( \nabla^\eta h^\lambda_\eta + \nabla^\eta h_\eta^\lambda - \nabla^\lambda h_{\alpha\beta}g^{\alpha\beta} \Big) \nabla_\lambda \alpha_{\mu\nu}  \nonumber \\
                                        &&\hspace{1.7cm} 
                                        + \Big( \nabla^\eta h^\lambda_\mu  + \nabla_\mu h^{\eta\lambda} - \nabla^\lambda h^\eta_\mu \Big) \nabla_\eta \alpha_{\lambda\nu} 
                                        + \Big( \nabla^\eta h^\lambda_\nu  + \nabla_\nu h^{\eta\lambda} - \nabla^\lambda h^\eta_\nu \Big) \nabla_\eta \alpha_{\lambda\mu} \Bigg).
 \ee
In following computations as a tensor $\alpha_{\mu\nu}$ we will use $\Box^n R_{\mu\nu}$.

The variation of the first two terms in \eqref{appaction} gives us the well known Einstein tensor plus a cosmological constant term:
\be
    \delta \int\! d^Dx \sqrt{|g|} \Big(R - 2 \Lambda_{\rm cc} \Big) = - \int\! d^Dx \sqrt{|g|} \Big(G^{\mu\nu} + \Lambda_{\rm cc} g^{\mu\nu} \Big) h_{\mu\nu}.
\ee
The variation of the next two terms are similar, but more complicated:
\be
&&    \boxed{\delta \! \int\! d^Dx \sqrt{|g|} (R \Box^n R) } \label{YD-J} \\
&& = \int\! d^Dx \Big( \delta\sqrt{|g|}R\Box^n R + 2 \sqrt{|g|} \delta R \Box^n R + \sqrt{|g|} \sum_{i=0}^{n-1} R \Box^{n-1-i} (\delta\Box) \Box^i R \Big) \nonumber \\
                 &&                          = \int\! d^Dx \sqrt{|g|} \Big( \half g^{\mu\nu} R \Box^n R - 2R^{\mu\nu}\Box^n R - 2g^{\mu\nu} \Box^{n+1} R + 2 \nabla^{\mu}\nabla^{\nu}\Box^n R
                 \nonumber \\
    && + \sum_{i=0}^{n-1} \nabla^{(\nu}\((\nabla^{\mu)}\Box^i R)( \Box^{n-1-i} R)\)
                                             - \half g^{\mu\nu} \sum_{i=0}^{n-1} \nabla^\sigma\((\nabla_\sigma\Box^i R)( \Box^{n-1-i} R)\) - \sum_{i=0}^{n-1} (\nabla^\mu\nabla^\nu\Box^i R)( \Box^{n-1-i} R) \Big) h_{\mu\nu}
    \nonumber \\
      && = \int\! d^Dx \sqrt{|g|} \Big( \half g^{\mu\nu} R \Box^n R - 2R^{\mu\nu}\Box^n R - 2g^{\mu\nu} \Box^{n+1} R + 2 \nabla^{\mu}\nabla^{\nu}\Box^n R
      \nonumber \\
    &&
    + \sum_{i=0}^{n-1} (\nabla^{(\mu}\Box^i R) ( \nabla^{\nu)}\Box^{n-1-i} R)
                                                    - \half g^{\mu\nu} \sum_{i=0}^{n-1} \nabla^\sigma\((\nabla_\sigma\Box^i R)(\Box^{n-1-i} R)\) \Big) h_{\mu\nu}\equiv\boxed{    \int\! d^Dx \sqrt{|g|} J^{\mu\nu} h_{\mu\nu} }\, ,\nonumber 
\ee

\be
&&  \hspace{-0.2cm}
 \boxed{ \delta \!  \int\! d^Dx \sqrt{|g|} (R^{\mu\nu} \Box^n R_{\mu\nu}) } \label{YD-K}
 \\
&& \hspace{-0.2cm}
= \int\! d^Dx \Bigg\{ \delta\sqrt{|g|} (R^{\mu\nu} \Box^n R_{\mu\nu}) + 2\sqrt{|g|} \delta g^{\rho\mu} g^{\sigma\nu} (R_{\mu\nu} \Box^n R_{\rho\sigma})  \nonumber \\
    && \hspace{-0.2cm}
    + 2 \sqrt{|g|} \delta R_{\mu\nu} \Box^n R^{\mu\nu} + \sqrt{|g|} \sum_{i=0}^{n-1} R^{\mu\nu} \Box^{n-1-i} (\delta \Box) (\Box^i R_{\mu\nu} ) \Bigg\}
   \nonumber \\
                                             &&    \hspace{-0.2cm}               = \int\! d^Dx \sqrt{|g|} \Bigg \{ \half g^{\mu\nu} R_{\alpha\beta} \Box^n R^{\alpha\beta} - 2 R^{\beta(\nu} \Box^n R^{\mu)}_{\ \ \beta}
     - \Box^{n+1} R^{\mu\nu} + 2 \nabla^\beta\nabla^{(\nu}\Box^n R^{\mu)}_{\ \ \beta} \nonumber \\
     && \hspace{-0.2cm}
     - g^{\mu\nu}\nabla_\alpha \nabla_\beta \Box^n R^{\alpha\beta}
     - 2 \sum_{i=0}^{n-1} \nabla_\beta \(( \Box^i R_\alpha^{\ (\nu} )(\nabla^{\mu)} \Box^{n-1-i} R^{\alpha\beta}) \)
   + 2 \sum_{i=0}^{n-1} \nabla_\beta \( (\Box^i R_\alpha^{\ \beta})( \nabla^{(\mu} \Box^{n-1-i} R^{\nu)\alpha}) \)
     \nonumber \\
    &&  \hspace{-0.2cm}
    +   \sum_{i=0}^{n-1} (\nabla^{(\nu}\Box^i R_{\alpha\beta})(\nabla^{\mu)} \Box^{n-1-i} R^{\alpha\beta})
    - \half g^{\mu\nu} \sum_{i=0}^{n-1} \nabla_\rho \( (\nabla^\rho \Box^i R_{\alpha\beta})( \Box^{n-1-i} R^{\alpha\beta}) \) \Bigg \} h_{\mu\nu}
      \!   \equiv \!\!  \int \! d^Dx \sqrt{|g|} K^{\mu\nu} h_{\mu\nu} .
       \nonumber
        \ee
Hence the EOM is:
\be \boxed{\boxed{
    E^{\mu\nu} = - 2 \kappa^{-2} \Big( G^{\mu\nu} + \Lambda_{\rm cc} g^{\mu\nu} + \omega_{\rm R} J^{\mu\nu} + \omega_{\rm Ric} K^{\mu\nu} \Big) = 0}}\,  .
    \label{YD-EOM}
\ee

\subsection{EOM with two form factors}
It is also interesting to see the compact form of EOM for the case of a theory with two form factors
\be
&& 
\hspace{-.5cm} \mathcal{L}_{\rm g} = -  2 \kappa_{D}^{-2} \, \sqrt{|g|} 
\Big[\Lambda_{\rm cc} + {\bf R} 
+
{\bf R} \, 
 \gamma_0(\Box
)
 {\bf R }
 + {\bf Ric} \, 
\gamma_2(\Box
)
 {\bf Ric} 
\Big]. 
\ee
For simplicity we reduced the term ${\bf Riem}\,\gamma_4(\Box){\bf Riem}$, which appeared in \eqref{gravity}, to two possible structures with ${\bf R}$ and ${\bf Ric}$ tensors respectively. This was possible under the sign of integral by exploiting Bianchi identities for ${\bf Riem}$ tensor and integration by parts - the result differs only by terms higher than quadratic in curvature (vertices), which are put in neglected here curvature potential ${\bf V}$. Each of the function $\gamma_0$, $\gamma_2$ has expansion as it was written in \eqref{formfactor}.
The form of the equation of motion is the following
\be
&& \hspace{-0.7cm}
E_{\mu\nu} 
 = \frac{ \delta \left[  \sqrt{|g|} \left( R - 2 \Lambda_{\rm cc}+ R 
\gamma_0(\Box) R+R_{\alpha \beta} 
\gamma_2(\Box) R^{\alpha \beta} \right) \right]}{\sqrt{|g|} \delta g^{\mu\nu}} \nonumber \\
 && = G_{\mu\nu} +  \Lambda_{\rm cc} \, g_{\mu \nu} -  
 \frac{1}{2} g_{\mu\nu} \left(R \gamma_0(\Box) R \right) -  
 \frac{1}{2} g_{\mu\nu} \left(R_{\alpha \beta} \gamma_2(\Box)
 R^{\alpha \beta} \right) 
 \nonumber \\
&&+   2\frac{\delta R}{\delta g^{\mu \nu}  } \left( \gamma_0(\Box)
R \right)+   
\frac{\delta R_{\alpha \beta}}{\delta g^{\mu \nu}  } \left( \gamma_2(\Box)
 R^{\alpha \beta} \right)+   \frac{\delta R^{\alpha \beta}}{\delta g^{\mu \nu}  } \left( \gamma_2(\Box)
 R_{\alpha \beta} \right)  \nonumber \\
 &&
 +  \frac{\delta \Box^r}{\delta g^{\mu\nu} }
 \left( 
\frac{  \gamma_0(\Box^l)
 -\gamma_0(\Box^r) 
}{\Box^r - \Box^l} 
 R R \right) 
+  \frac{\delta \Box^r}{\delta g^{\mu\nu} }
 \left( 
  \frac{ \gamma_2(\Box^l)
- \gamma_2(\Box^r)
}{\Box^r - \Box^l} 
 R_{\alpha \beta} R^{\alpha \beta} \right) \, ,
\ee
where $\Box^{l,r}$ act on the left and right arguments (on the right of the incremental ratio) as indicated inside the brackets.
It is of great importance for finding approximate solutions to study the expansion of this EOM in powers of gravitational curvatures \cite{Yiwei}. Here we will concentrate mainly on an approximation, which retains only terms linear in curvature. In this case EOM greatly simplifies to
\be
 \hspace{-0.8cm} E_{\mu\nu}  =  G_{\mu\nu} + \Lambda_{\rm cc} \, g_{\mu \nu}  
 +   \! 2\frac{\delta R}{\delta g^{\mu \nu}  }  \gamma_0(\Box)
R +   
\frac{\delta R_{\alpha \beta}}{\delta g^{\mu \nu}  }  \gamma_2(\Box)
 R^{\alpha \beta} 
 +   \frac{\delta R^{\alpha \beta}}{\delta g^{\mu \nu}  }  \gamma_2(\Box)
 R_{\alpha \beta}  + O({\cal R}^2).
\ee
To this order we can put simplified variations $\delta R=g^{\alpha\beta}\delta R_{\alpha\beta}$ and $\delta R^{\alpha\beta}=g^{\alpha\kappa}g^{\beta\lambda}\delta R_{\kappa\lambda}$ and get
\be
 && E_{\mu\nu} = G_{\mu\nu} + \Lambda_{\rm cc} \, g_{\mu \nu}  
 +   2\frac{\delta R_{\alpha\beta}}{\delta g^{\mu \nu}  } \left( g^{\alpha\beta}\gamma_0(\Box)
R +\gamma_2(\Box)
 R^{\alpha \beta}\right) +O({\cal R}^2)\,.
\ee
We see, that if the first condition from \eqref{formfactor} holds ($\gamma_2=-2\gamma_0$), then this simplifies even more and we get gravitational EOM, when coupled to matter
\be
 &&  G_{\mu\nu} + \Lambda_{\rm cc} \, g_{\mu \nu}  
 +   2\frac{\delta R_{\alpha\beta}}{\delta g^{\mu \nu}  } \gamma_2(\Box)
 G^{\alpha \beta} +O({\cal R}^2)=T_{\mu\nu}\,.
\ee
Using the fact that under the integral 
\be
\frac{\delta R_{\alpha\beta}}{\delta g^{\mu \nu}  }=\frac{1}{2}g_{\alpha(\mu}g_{\nu)\beta}\Box+\frac{1}{2}g_{\mu\nu}\nabla_\alpha
\nabla_\beta-g_{\alpha(\mu|}\nabla_\beta\nabla_{|\nu)} \, , 
\ee
 and contracted Bianchi identity $\nabla^\mu G_{\mu\nu}=0$ we arrive to EOM in a form
\be
 &&  G_{\mu\nu} + \Lambda_{\rm cc} \, g_{\mu \nu}  
 +   \Box \gamma_2(\Box)
 G_{\mu \nu} +O({\cal R}^2)=T_{\mu\nu}\,.
\ee
Reminding from \eqref{formfactor} that $\gamma_2(\Box)=\frac{e^{H(-\Box_\Lambda)}-1}{\Box}$
we rewrite above as
\be
 &&    \Lambda_{\rm cc} \, g_{\mu \nu}  
 +  e^{H(-\Box_\Lambda)}
 G_{\mu \nu} +O({\cal R}^2)=T_{\mu\nu}\,.
\ee
Finally we define effective energy tensor of the system as
\be
T_{{\rm eff},\,\mu\nu}=e^{-H(-\Box_\Lambda)}\(T_{\mu\nu} -  \Lambda_{\rm cc} \, g_{\mu \nu}\)=e^{-H(-\Box_\Lambda)}T_{\mu\nu} - e^{-H(0)} \Lambda_{\rm cc} \, g_{\mu \nu} ,
\ee
and cast the equation in the familiar Einstein form
\be
 &&
 G_{\mu \nu} =T_{{\rm eff},\,\mu\nu}+O({\cal R}^2)\,.
\ee
This EOM is divergence-free thanks to the Bianchi identity of the Einstein tensor. The presence of the cosmological constant term does not spoil the analysis here, and for the standard choice $H(0)=0$ it does not get delocalized in effective energy-momentum tensor of the system.

In the case, where there is no relation between two form factors $\gamma_0(\Box)$ and $\gamma_2(\Box)$, the full EOM takes the following form
\be
 &&  G_{\mu\nu} + \Lambda_{\rm cc} \, g_{\mu \nu}  
+\Box\(g_{\mu\nu}\gamma_0(\Box)R+\gamma_2(\Box)R_{\mu\nu}\)
+g_{\mu\nu}\nabla^\alpha\nabla^\beta\(g_{\alpha\beta}\gamma_0(\Box)R
+\gamma_2(\Box)R_{\alpha\beta}\)\nonumber\\
&&-2\nabla^\alpha\nabla_{(\mu}\(g_{\nu)\alpha}\gamma_0(\Box)R
+\gamma_2(\Box)R_{\nu)\alpha}\)+O({\cal R}^2)=T_{\mu\nu}\,.
\ee
Here the simplification with the first term, Einstein tensor, does not happen and effective EOM can not be written in the Einstein form for some effective energy tensor. Using contracted Bianchi identity $\nabla^\mu R_{\mu\nu}=1/2\nabla_\nu R$ and neglecting additional curvature tensors resulting from commutation of covariant derivatives we write above EOM as
\be
&&G_{\mu\nu} +  \Lambda_{\rm cc} \, g_{\mu \nu}  
+\Box\left(g_{\mu\nu}\left(2\gamma_{0}(\Box)+\frac{1}{2}\gamma_{2}(\Box)\right)R
+\gamma_{2}(\Box)R_{\mu\nu}\right)\nonumber\\
&&-\nabla_{\mu}\nabla_{\nu}\left(2\gamma_{0}(\Box)+\gamma_{2}(\Box)\right)R
+O({\cal R}^2)=T_{\mu\nu}\,.
\ee
Now we reshuffle some terms and use the second condition in \eqref{formfactor}, and eventually EOM takes the simple form
\be
&&e^{H_{2}(-\Box_{\Lambda})}G_{\mu\nu} + \Lambda_{\rm cc} \, g_{\mu \nu}+\left(g_{\mu\nu}\Box-\nabla_{\mu}\nabla_{\nu}\right)\left(2\gamma_{0}(\Box)
+\gamma_{2}(\Box)\right)R+O({\cal R}^2)=T_{\mu\nu}\,.
\label{EOMtwoff}
\ee
As we see the structure of terms is different than in the case of one form factor. However, this is still a good gravitational EOM, because is divergence-free. Here this feature is brought by the fact of transversality in $\mu,\nu$ indices of the first variation of Ricci tensor. We have namely that
\be
\nabla^\mu \frac{R_{\alpha\beta}}{\delta g^{\mu\nu}}=0\,,
\ee
which can be also understood from the fact of the presence of terms proportional to the transverse projector $g_{\mu\nu}\Box-\nabla_{\mu}\nabla_{\nu}$ in \eqref{EOMtwoff}.

Both theories (with one or two independent form factors) are gauge theories of metric fluctuations and they can be coupled consistently to conserved matter energy tensor. We notice that in the second case the IR limit ($z\to 0$) and the limit of small curvature do not coincide. For the case of entire functions  expanding to the first order in $z$ we get $2\gamma_{0}(\Box)+\gamma_{2}(\Box)\sim 1/\Lambda^2$ (because of dimensional reasons), and in 
\eqref{EOMtwoff} the contributions $\nabla^2 R$ are suppressed.


\end{document}